\documentclass{jfm}
\usepackage{graphicx}
\usepackage{natbib}
\usepackage[normalem]{ulem}
\usepackage{soul}
\usepackage{amssymb,amsmath}
\usepackage{color}
\usepackage{cases}
\usepackage{wasysym}
\usepackage{easybmat}   

\ifCUPmtlplainloaded \else
  \checkfont{eurm10}
  \iffontfound
    \IfFileExists{upmath.sty}
      {\typeout{^^JFound AMS Euler Roman fonts on the system,
                   using the 'upmath' package.^^J}%
       \usepackage{upmath}}
      {\typeout{^^JFound AMS Euler Roman fonts on the system, but you
                   dont seem to have the}%
       \typeout{'upmath' package installed. JFM.cls can take advantage
                 of these fonts,^^Jif you use 'upmath' package.^^J}%
      }
  \else
  \fi
\fi

\ifCUPmtlplainloaded \else
  \checkfont{msam10}
  \iffontfound
    \IfFileExists{amssymb.sty}
      {\typeout{^^JFound AMS Symbol fonts on the system, using the
                'amssymb' package.^^J}%
       \usepackage{amssymb}%

      }{}
  \fi
\fi

\ifCUPmtlplainloaded \else
  \IfFileExists{amsbsy.sty}
    {\typeout{^^JFound the 'amsbsy' package on the system, using it.^^J}%
     \usepackage{amsbsy}}
    {}
\fi
\usepackage{ifsym}

\newsavebox{\astrutbox}
\sbox{\astrutbox}{\rule[-5pt]{0pt}{20pt}}

\newcount\ndots
\def\drwln#1#2{\raise 2.5pt\vbox{\hrule width #1pt height #2pt}}

\def\solid{\drwln{24}{1}}

\newcommand{\DoubleInfSum}[0]{\displaystyle\sum_{k=-\infty}^{+\infty}}

\newcommand{\expiomega}{e^{i\,\omega\;t}}

\title[Modeling of a Thermoacoustic Heat Engine]{Linear and Nonlinear Modeling of a Traveling-Wave Thermoacoustic Heat Engine}

\author[Carlo Scalo, Sanjiva K. Lele, Lambertus Hesselink] %
{C\ls A\ls R\ls L\ls O\ls \ns S\ls C\ls A\ls L\ls O\ls$^1$%
  \thanks{Email address for correspondence: scalo@stanford.edu},\ns
S\ls A\ls N\ls J\ls I\ls V\ls A\ls \ns K.\ns L\ls E\ls L\ls E\ls$^2$\break
L\ls A\ls M\ls B\ls E\ls R\ls T\ls U\ls S\ls \ns H\ls E\ls S\ls S\ls E\ls L\ls I\ls N\ls K\ls $^3$\break
}

\affiliation{$^1$Center for Turbulence Research, Stanford, CA, 94305, USA \\[\affilskip]
$^2$Dept. of Aeronautics and Astronautics and Mechanical Eng., Stanford, CA, 94305, USA \\[\affilskip] $^3$Dept. of Aeronautics and Astronautics and  Electrical Eng., Stanford, CA, 94305, USA}

\pubyear{2014}
\volume{XXX}
\pagerange{XX--XX}
\date{?; revised ?; accepted ?. - To be entered by editorial office}
\begin{document}

\maketitle

\begin{abstract}

%%% What have we done, motivation, computational setup.. 
We have carried out three-dimensional Navier-Stokes simulations, from quiescent conditions to the limit cycle, of a theoretical traveling-wave thermoacoustic heat engine (TAE) composed of a long variable-area resonator shrouding a smaller annular tube, which encloses the hot (HHX) and ambient (AHX) heat-exchangers, and the regenerator (REG). Simulations are wall-resolved, with no-slip and adiabatic conditions enforced at all boundaries, while the heat transfer and drag due to the REG and HXs are modeled. HHX temperatures have been investigated in the range 440K -- 500K with AHX temperature fixed at 300K.
 % A secondary ambient heat exchanger (AHX2) is introduced to achieve a limit cycle, reproducing equilibrium conditions typically found in thermal buffer tubes (TBT).
%%% What have we discovered.. 
% 1) linear regime 
The initial exponential growth of acoustic energy is due to a network of traveling waves thermoacoustically amplified by looping around the REG/HX unit in the direction of the imposed temperature gradient. A simple analytical model demonstrates that such instability is a localized Lagrangian thermodynamic process resembling a Stirling cycle. An inviscid system-wide linear stability model based on Rott's theory is able to accurately predict the operating frequency and the growth rate, exhibiting properties consistent with a supercritical Hopf bifurcation.
% 2) nonlinear regime, towards saturation
The limit cycle is governed by acoustic streaming -- a rectified steady flow resulting from high-amplitude nonlinear acoustics. Its key features are explained with an axially symmetric incompressible model driven by the wave-induced stresses extracted from the compressible calculations. These features include Gedeon streaming, Rayleigh streaming in the resonator, and mean recirculations due to flow separation. The first drives the mean advection of hot fluid from the HHX to a secondary heat exchanger (AHX2), in the thermal buffer tube (TBT), necessary to achieve saturation of the acoustic energy growth. The direct evaluation of the nonlinear energy fluxes reveals that the efficiency of the device deteriorates with the drive ratio and that the acoustic power in the TBT is balanced primarily by the mean advection and thermoacoustic heat transport.
\end{abstract}


\section{Introduction}

% [INTRO/BRIEF HISTORY] What are TAE, why are they interesting (motivation), very brief technological history..
Thermoacoustic heat engines (TAE) are devices that can convert available thermal energy into acoustic power with very high
efficiencies. This potential is due to the absence of moving parts and relative simplicity of the components. This
results in low manufacturing and maintenance costs making these systems an attractive alternative for clean and
effective energy generation or waste-energy reutilization. The core energy conversion process occurs in the regenerator -- a porous metallic block,
placed between a hot and a cold heat-exchanger, sustaining a mean temperature gradient in the axial direction. Acoustic
waves propagating through it (under the right conditions) can be amplified via a thermodynamic process resembling a Stirling
cycle. Most designs explored up to the mid 1980's were based on standing waves and had efficiencies typically
less than 5\%. A significant breakthrough was made by \cite{Ceperley_1979_JAcoustSA} who showed that traveling-waves can
extract acoustic energy more efficiently, leading to the design concept for traveling-wave TAEs
\citep{BackhausS_JAcoustSocAm_2000,deWaele_2009_JSV}. In this configuration the generated acoustic power is in part 
resupplied to the regenerator via some form of feedback and in part directed towards a resonator for energy extraction. A secondary
ambient heat exchanger is typically needed to contain the heat leaking (due to nonlinear effects) from the hot-heat exchanger.
This design is the focus of the present study.

%% Motivation
Improving the technology behind TAEs has been of particular interest in the last decade with research efforts being made
worldwide (see \cite{Garrett_2004_AmJPhys} for a review). A recent breakthrough, for example, has been made by \cite{TijaniS_2011_JApplPhys}
who designed a traveling-wave TAE achieving a remarkable overall efficiency of 49\% of the Carnot limit. The state-of-the-art prediction capabilities and technological design of TAEs can, however, significantly benefit from a high-fidelity description of the underlying fluid mechanics. The potential of such approach to fill important modeling gaps has inspired the present study which relies on full-scale three-dimensional flow simulations to gain insight into the linear and nonlinear processes occuring in a theoretical traveling-wave thermoacoustic engine.

% [MODELING] %%% Theoretical modeling (from Rott to Swift), anticipation of problems of state of the art modeling and open question
A comprehensive theoretical analysis of thermoacoustic effects in ducts is provided in the seminal publications by Rott and co-workers \citep{Rott_ZAMP_1969,Rott_ZAMP_1973,Rott_ZAMP_1974,Rott_ZAMP_1975,Rott_ZAMP_1976,Rott_NZZ_1976,ZouzoulasR_ZAMP_1976,Rott_1980_AdvApplMech,MullerR_ZAMP_1983,Rott_JFM_1984} where a predictive analytical framework (restricted to simple configurations) is derived improving upon pre-existing theories by \cite{Kirchhoff_PoggAnn_1868} and \cite{Kramers_Physica_1949}. Issues addressed include the onset of instability, thermoacoustic heating, transport due to acoustic nonlinearities and effects of variable cross-sectional area. Later, Swift and co-workers used Rott's work as the basis for the development
of semi-empirical low-order models for the acoustics in various components found in real thermoacoustic engines \citep{Swift_1988_JAcoustSA,Swift_JAcoustSocAm_1992,SwiftS_JThermPhysHeatTr_1990}. This resulted in the development of the prediction software
package {\sc DeltaE} \citep{WardS_JAcouSocAm_1994} (replaced now by {\sc DeltaEC}), which, together with similar modeling tools such as {\sc SAGE} and {\sc REGEN},
is still actively used in the academic literature as well as in industry. Other examples of advanced low-order modeling relying on Rott's theory and
systematic asymptotic approximations \citep{Bauwens_JFM_1996,PanhuisRMS_JFM_2009,HirecheWCQFB_CRM_2010} suffer from similar shortcomings.
While the prediction of global quantities of interest such as acoustic amplitude, efficiency, and frequency of operation can be made accurate 
in low-pressure amplitude cases (not without some heuristics required on the user's end), it is not possible with such an approach to directly account, for example, for the interaction of high-amplitude acoustic wave with complex geometries, the effects of transitional turbulence and higher-order harmonics on thermoacoustic transport \citep{OlsonS_Cryo_1997} and acoustic energy dissipation \citep{WardS_JAcouSocAm_1994}. First-principles modeling tools (e.g. direct numerical computations) can successfully address such issues, which have a direct impact on the functionality of TAEs and are, nonetheless, either (inevitably) ignored or heuristically treated.

%% [ACOUSTIC nonlinearITIES]
The most important nonlinear process impacting the efficiency of TAEs is acoustic streaming \citep{BoluriaanM_2003_IntlJournalAero}.
This is the cumulative effect of fluid parcel displacements over several high-amplitude acoustic cycles. The result is a
rectified flow that, when unsteady, may evolve over time scales orders of magnitude larger than the fundamental acoustic frequency \citep{ThompsonAM_2004_JAcoustSocAm_2}. In TAEs, streaming is responsible for mean advection of hot fluid away from the HHX (thermal leakage)
and limiting the obtainable wave amplitude. Penelet and co-workers \citep{PeneletGLB_PhysLett_2006,PeneletGLB_PhysRevE_2005,PeneletJGLB_ActaAcust_2005,PeneletGGD_IntJHMT_2012} identified in the inadequate modeling of
such nonlinear effects the primary reason for the failure of low-order models to correctly capture wave-amplitude saturation, even in simple geometries.
It is therefore necessary to adopt a direct modeling approach as done by \cite{BoluriaanM_2003_AIAA} who performed simulations solving the fully
compressible viscous flow equations in an idealized two-dimensional configuration modeling traveling-wave streaming suppressed by a jet pump.
To properly account for the viscous interactions with the solid wall, the Stokes thickness needs to be resolved. Three-dimensional flow simulations
of similar configurations, fully resolving thermo-viscous effects and transport due to streaming, are yet to be attempted. In the present work we
take on the challenge of studying streaming occurring in a three-dimensional flow with geometric complexities by analyzing its effects on the engine performance but also directly modeling it with a vorticity-streamfunction formulation. A fluid dynamic analogy can be exploited to model the streaming \citep{RudenkoS_SSS_1977}
as an incompressible flow driven by wave-induced Reynolds stresses. By following this approach, we developed a simplified numerical model to gain
insight into the spatial structure of the acoustic stresses and their relationship with complex geometrical features. The model reproduces the key nonlinear effects such as Rayleigh and Gedeon streaming \citep{Gedeon_1997_Cryocoolers}.

% [CHALLENGES TO FULL-SCALE MODELS]: 1) boundary layer thicknesses 2) incompressible/compressible coupling of REG with resonator
A very high computational cost, however, is associated with the direct resolution of the governing equations. In a full TAE
the range of temporal and spatial scales can span 4 orders of magnitude \citep{HamiltonIZ_JAcoustSocAm_2002}. These range from the acoustically
driven thermal and viscous boundary layers, scaling with the Stokes boundary layer thickness $\delta_\nu$ typically of the order of $10^{-1}$ mm,
to the resonator length, typically of the order of the acoustic wavelength $\lambda \simeq 1$ m. This challenge has been directly confronted in
the present simulations where special care has been taken to devise a meshing strategy that could capture the large range of scales while
retaining a manageable computational cost in three-dimensions. The porous metallic structure of the regenerator and heat-exchangers has not been directly resolved in our work due to its complex geometrical features. Directly resolving such structures, however, does not pose a significant extra burden on the required computational time per se (since the characteristic pore size is typically 100-500 microns, of the order of the already resolved viscous boundary layer) but, rather, on the meshing effort, which would become unfeasible, and on the modeling side, requiring to account for conduction through the metallic structure and coupling with the fluid.

% [OUR CONTRIBUTION]: say something like, we hope to lay the groundwork for full-scale, time-resolved Navier-Stokes simulations
For the purpose of the present study we adopt a theoretical model of a traveling-wave thermoacoustic heat-engine, building upon the design proposed by \cite{NijeholtTS_2005_JAcoustSA}, extending it to a three-dimensional setup and adding a secondary ambient heat-exchanger necessary to achieve a limit cycle. The heat-exchangers and regenerators are modeled using semi-empirical source terms available in the literature. This allows us to 
focus on the full-scale resolution of the high-amplitude acoustics interacting with complex geometrical features. Details omitted in \cite{NijeholtTS_2005_JAcoustSA}
regarding the geometry and the modeling of the heat-exchangers and regenerators are reconstructed to the best of the authors' ability with the help of Ray Hixon (pers. comm., 2012). The present work can be regarded as the first step towards a simple benchmark case for validation of computational modeling of thermoacoustic devices. The spatially and temporally
resolved data will be used to gain insight into the aforementioned linear and nonlinear governing physical processes and contribute to their modeling. The lack of experimental data available for the proposed theoretical device has required the adoption of several companion lower-order models to verify the results obtained from the three-dimensional simulations. The levels of modeling adopted range from zero-dimensional and purely analytical to axially symmetric and nonlinear.

% [PAPER OUTLINE]
In the following, the computational setup is first introduced, discussing the adopted meshing strategy and the semi-empirical heat-transfer and
drag models for the heat-exchangers and the regenerator. Results follow investigating first the start-up phase, where linear models are
adopted to describe the nature of the wave propagation throughout the system and amplification via thermoacoustic instability. Instantaneous
data is then collected at the limit-cycle where acoustic streaming is investigated. Results from an incompressible numerical model used to directly solver for streaming flow are discussed. Finally, thermal energy budgets in the TBT are analyzed.

%\Carlo{Also, look into recent work on Thermoacoustic engines published in JFM by a ph.d. student of Reinstra W. on asymptotic treatment of sounds vs porous material. If not just to mention it in the literature review. I think this is \cite{PanhuisRMS_JFM_2009}}
% {\bf Note that \cite{YuLDH_JApplPhys_2007} seem to have done a pretty good job, but it's 2D and it's Fluent. They cite experimental data but the guys might be cheating.}
%\Carlo{Make sure you cite T. Sujith (several JFM papers) and M. Juniper (Cambridge) which have stressed the importance of non-modal, i.e. transient growth, in the context of thermoacoustics.} 

\section{Problem formulation} \label{sec:problem_formulation} % governing equations, CFD model and boundary conditions

\begin{figure}
\centering
 \includegraphics[width=\linewidth]{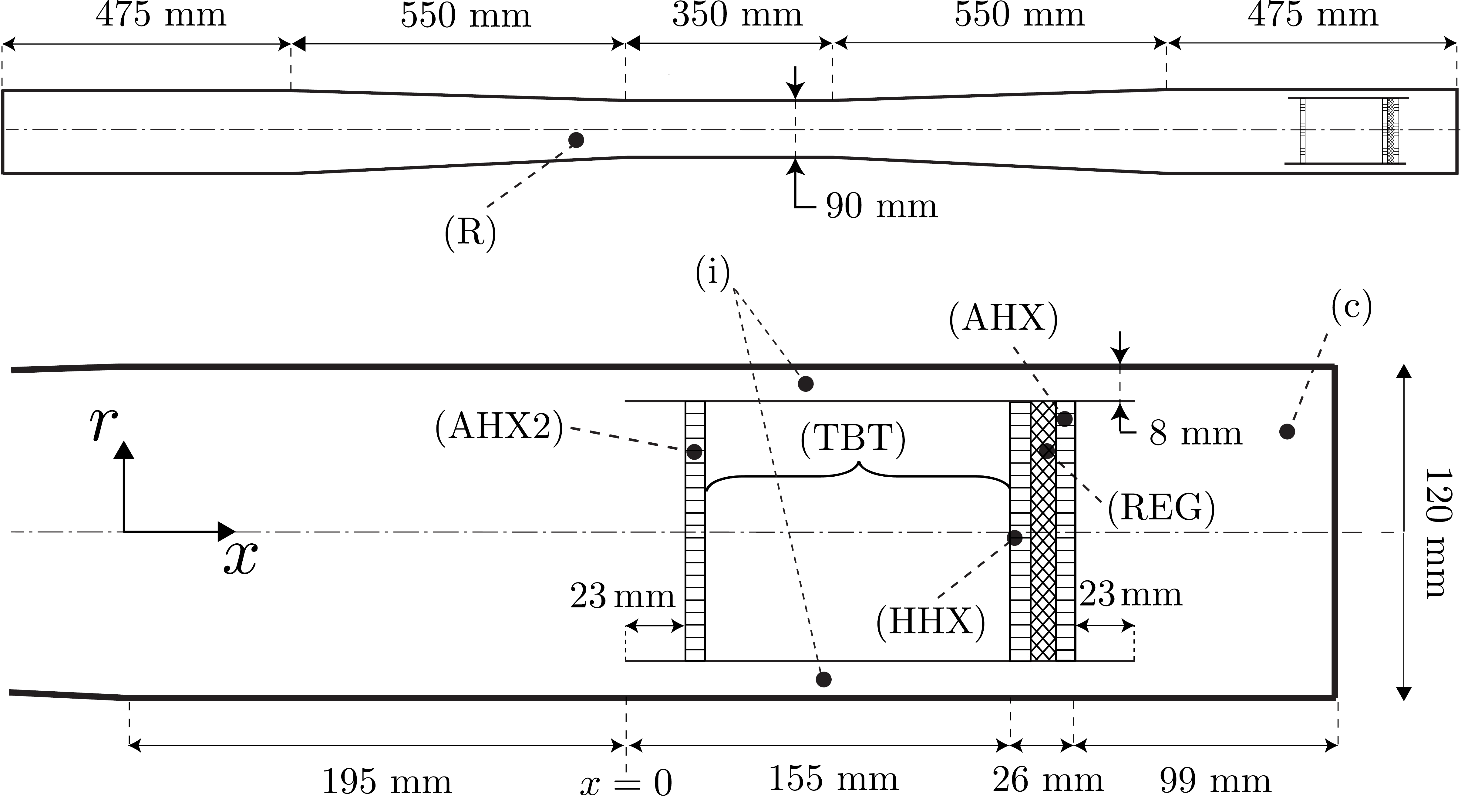}
 \caption{Illustration of the model traveling-wave thermoacoustic engine inspired by {\protect \cite{NijeholtTS_2005_JAcoustSA}}.
 Full system (top), regenerator/heat-exchanger area (bottom), drawn to scale. A dash-dotted line indicates the line of symmetry.
 The different components of the engine are the resonator (R), the annular feedback inertance loop (i), 
 the compliance (c), the hot heat exchanger (HHX), the regenerator (REG), the ambient heat exchanger (AHX), secondary heat exchanger (AHX2), and the thermal buffer tube (TBT). The annular tube enclosing the AHX2, HHX, REG, and AHX is also refered to as pulse tube, of which the TBT is only a section.}
 \label{fig:computationalsetupNijeholt}
\end{figure}

The full-scale approach advocated in the present study requires the resolution of the complete set of compressible viscous flow equations. The conservation equations for mass, momentum and total energy are reported here in index notation
\begin{subnumcases}{\label{eq:governingEquations}}
 \frac{\partial \rho}{\partial t} + \frac{\partial \rho u_i}{\partial x_i} = 0 \\
\frac{\partial \rho u_i}{\partial t} + \frac{\partial \rho u_i u_j }{\partial x_j} + \frac{\partial \tau_{ij}}{\partial x_j} = -\frac{\partial p}{\partial x_i}  + D_i \\
\frac{\partial \rho E_t}{\partial t} + \frac{\partial }{\partial x_j}\left[ \left( \rho E_t + p \right) u_j \right] +  \frac{\partial \beta_j}{\partial x_j}= S_E
 \end{subnumcases}
where $x_1$, $x_2$ and $x_3$ (or $x$, $y$ and $z$) are the axial and cross-sectional coordinates, $u_i$ the velocity components in those directions, $p$, $\rho$ and $E_t$ are the pressure, density and total energy. The viscous stress tensor and energy flux, respectively, $\tau_{ij}$ and $\beta_i$, are given by
\begin{subnumcases}{\label{eq:heat_and_viscousflux}}
\beta_i = -u_j \tau_{ij}  + q_i \\
\tau_{ij} = -2 \mu \left[ S_{ij} - \frac{1}{3}\frac{\partial u_k}{\partial x_k} \delta_{ij} \right]
 \end{subnumcases}
where $q_i$ is the molecular heat-flux, $\mu$ the dynamic viscosity, $S_{ij}$ the deviatoric part of the strain-rate tensor. The fluid is assumed to be an ideal gas with reference state given by $\rho_\textrm{ref} = 1.2$ kg/m$^3$, $p_\textrm{ref} = 101325$  Pa and $T_\textrm{ref} = 300$ K. The dynamic viscosity varies with temperature based on the power law $\mu(T) = \mu_\textrm{ref} \left( T / T_\textrm{ref} \right)^{0.76} $. The Prandtl number is $Pr = 0.7$ for all cases.
The equations are solved in the geometry illustrated in figure \ref{fig:computationalsetupNijeholt} consisting of a long resonator with the heat-exchanger/regenerator (REG/HX) unit at one end, enclosed by a zero-thickness annular tube, which is also referred to as the pulse tube. The section of the pulse tube between the HHX and the AHX2 is the thermal buffer tube (TBT). The resonator and annular tube are treated as adiabatic no-slip walls with homogeneous Neumann conditions for pressure.

The regenerator and heat-exchangers are typically composed of a porous metallic structure ranging from overlapped wire screens (or metal felts) to stacks of parallel plates or rods, the former being more typical for regenerators, the latter for heat-exchangers. Following \cite{NijeholtTS_2005_JAcoustSA} we choose to model the heat transfer and drag in such components via the source terms $D_i$ and $S_E$ on the right-hand side of (\ref{eq:governingEquations}b) and (\ref{eq:governingEquations}c). They are expressed as 
\begin{subnumcases}{\label{eq:sourceTerms}}
D_i = -\left[R_C + R_F (u_j u_j)^{1/2}\right] u_i \\
S_E  = - u_i D_i + S_h
\end{subnumcases}
where drag term $D_i$ is modeled following the parametrizations
\begin{subnumcases}{\label{eq:R_C_R_F}}
 R_C = C_{sf} \mu \frac{(1-\phi)}{4 {d_w}^2 \phi} \\ 
 R_F = \frac{\rho C_{fd}}{4 d_w}
\end{subnumcases}
where $\phi$ and $d_w$ are the characteristic porosity and mesh wire size of the component, and $C_{sf}$ and $C_{fd}$ are dimensionless fitting constants taken from the ILK Dresden and K\"uhl metal felts correlations \citep{ThomasP_AIAA_2000}. These are specific to TAE regenerators and are derived under different oscillating flow conditions. Unfortunately, \cite{NijeholtTS_2005_JAcoustSA} does not provide values for the wire size $d_w$, which have been estimated by using the correlation suggested by \cite{Organ_CUPress_1992}
\begin{equation} \label{eq:organ_correlation}
 r_h = \frac{ d_w \phi }{4 (1-\phi)},
\end{equation}
where $r_h$ is the pore hydraulic radius.

The source term $S_h$ in (\ref{eq:sourceTerms}b) accounts for heat-transfer in the REG/HX unit and is modeled as
\begin{equation} \label{eq:energySource_heat}
 S_h = -\alpha_T \big[ T-T_0(x) \big],
\end{equation}
where $T$ is the instantaneous fluid temperature and $T_0(x)$ is the target mean temperature profile which is equal to $T_h$ and $T_a$, respectively, in the hot and ambient heat exchangers, and varies linearly in the regenerator between the two values. No information is provided in \cite{NijeholtTS_2005_JAcoustSA} for the proportionality constant $\alpha$. We propose to model it as (Ray Hixon, pers. comm, 2012)
\begin{equation} \label{eq:alpha_T_definition}  % \ref{eq:R_C_R_F}a due to \cite{Organ_CUPress_1992} and \ref{eq:alpha_T_definition} \citep{BejanA_2004_HeatTransfer}
 \alpha_T = \alpha_h \frac{\rho R}{\gamma -1}
\end{equation}
where $R$ is the gas constant and $\gamma$ is the ratio of specific heats. The ratio $\rho R /(\gamma -1)$ in \eqref{eq:energySource_heat} is a ballpark estimate of variations of total energy with respect to temperature ($\partial \rho E_t / \partial T$) derived using the equation of state. The constant $\alpha_h$ is defined as
\begin{equation}
\alpha_h = 1/\tau_h
\end{equation}
where $\tau_h$ is the characteristic time scale for heat-transfer in the void spaces of the heat exchangers and regenerator. An estimate for $\tau_h$ can be derived by 
modeling such components as stacks of parallel plates with spacing $b$ matching the given hydraulic radius of the pores, $b = 2 r_h$ (table \ref{tbl:drag_parameters}). This results in \citep{BejanA_2004_HeatTransfer}
\begin{equation}
\tau_h = \frac{(b/2)^2}{k / \rho C_p} = \frac{b^2 \rho Pr}{4 \mu}
\end{equation}
where $k$ is the thermal conductivity, $C_p$ the specific heat capacity of the gas.

This simplified model is expected to predict the intensity of the heat transfer rate to the pore flow within an order of magnitude. It is based on the assumption of perfect thermal contact and is in quantitative agreement with a similar model used in {\sc DeltaEC} \citep{WardS_JAcouSocAm_1994}. While the thermal regime resulting from \eqref{eq:energySource_heat} is not affected by the flow velocity, its linear dependency from the temperature facilitates the lower-order modeling efforts made in the present manuscript. Overall, despite their simplicity and coarse approximation, the application of the source terms \eqref{eq:sourceTerms} reproduces the essential thermodynamic and hydrodynamic processes that occur in regenerators and heat-exchangers in TAEs, as discussed in the following.

\begin{table}
  \centering
  \caption{Parameters for regenerator/heat-exchanger model (\ref{eq:sourceTerms}). Hydraulic radius $r_h$, characteristic wire diameter $d_w$ (mm),  porosity $\phi$, and drag coefficients $C_{sf}$, $C_{fd}$ \citep{ThomasP_AIAA_2000}. Values for  $r_h$ and $\phi$ are taken from {\protect \cite{NijeholtTS_2005_JAcoustSA}}. }
  \label{tbl:drag_parameters}
  \begin{tabular}{c|rr}
      \hline
        Parameter & Heat Exchangers & Regenerator \\
       \hline    
       $r_h$(mm) & 0.1& 0.041 \\
       $d_w$(mm)  &  0.2667 & 0.0670 \\
       $\phi$ & 0.60 & 0.71 \\
       $C_{sf}$ & 49.46  & 49.46 \\ 
       $C_{fd}$ & 0.572  & 0.572 \\
  \end{tabular}
\end{table}

\section{Numerical Model}

\begin{figure}
  \centering
\includegraphics[keepaspectratio=true,width=0.95\linewidth]{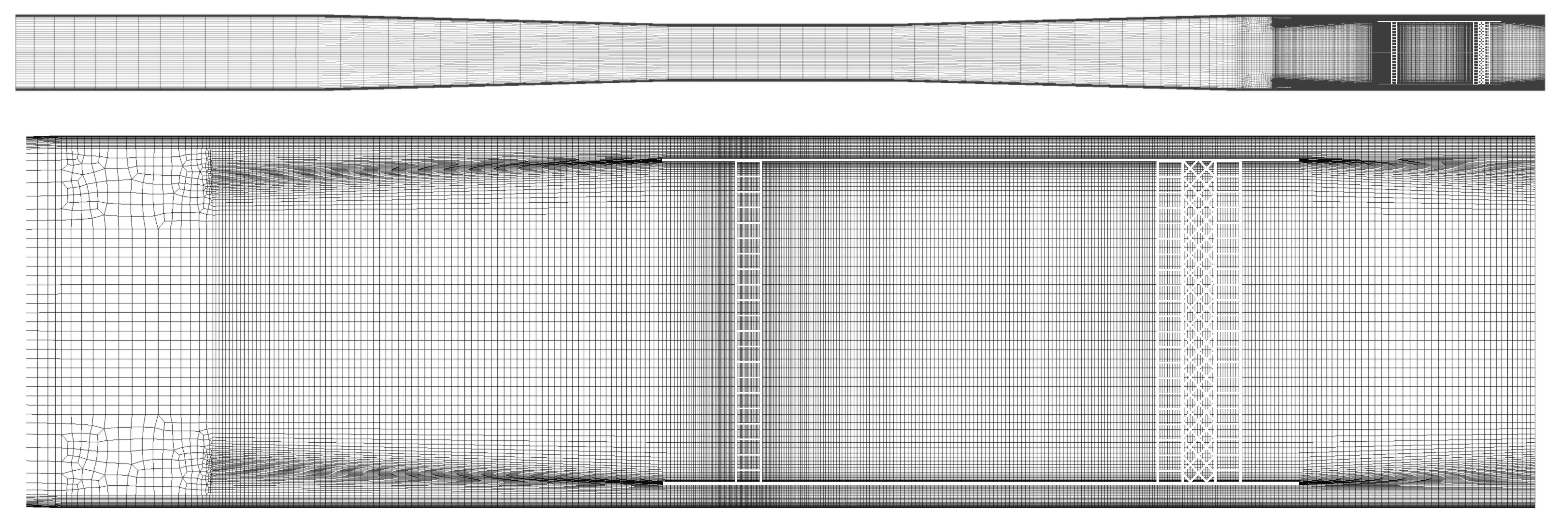}
\put(-377,95){(a)}
\put(-377,5){(b)}
 \caption{Cross-section of three-dimensional computational grid A for the full length resonator (a), zoom on the right end 
 (b) corresponding to view in figure \ref{fig:computationalsetupNijeholt}. The computational grid has been mirrored about
 the centerline for illustrative purposes. Computations are performed in a 90${^\circ{}}$ sector.}
\label{fig:computational_grid}
\end{figure}

% Describe mesh and geometry
The governing equations are discretized on an unstructured hexahedral mesh (figure \ref{fig:computational_grid})
and solved in a 90${^\circ{}}$ sector with rotational periodicity applied in the azimuthal direction. The infinitely
thin annular tube wall is introduced by breaking the mesh connectivity, creating two overlapping boundary surfaces
with opposite orientation. Three concentric \emph{O-grids} in the cross-section, two in the annular tube and one in
the pulse tube (not shown), are required in this region to map the polar mesh at the resonator walls to a quasi-uniform 
Cartesian block at the center. High resolution is retained near the sharp edges of the annular tube walls to properly 
capture the shear layer caused by periodic flow separation. Visual inspection of the flow in 
previous calculations does not reveal a significant vorticity magnitude away from the wall for $x<-0.1$ m. This has led
to the choice of collapsing hexahedral elements into larger ones (i.e. grid coarsening) starting at $x=$-0.146 m (figure \ref{fig:computational_grid}b), 
resulting in a coarser radial grid distribution for $x<$-0.146 m. Points have also been concentrated
in the AHX2 ($x$ = 0.031 m), due to intense instantaneous temperature and velocity gradients at the limit cycle created by hot fluid streaming away from the HHX (discussed later). Overall, a significant effort has been made to retain a high-quality structured grid when possible.

% How we decide on the resolution (weak claim of grid convergence)
A preliminary grid refinement study has been carried out to ensure adequate resolution of the axially symmetric components 
of the acoustic field and accurate prediction of the growth rate. The latter is sensitive primarily to the resolution in 
the axial direction (main direction of propagation of the acoustic waves), both in the resonator and in the 
thermal buffer tube. The viscous boundary layers are resolved with resolution of 0.1 mm at the wall for all cases. These considerations have lead to the
design of a baseline grid distribution, grid A, (figure \ref{fig:computational_grid}) used to rapidly advance in time 
through the initial transient. Simulations have been carried out for hot-heat exchanger temperatures of $T_h$ = 440K, 460K, 480K and 500K and ambient heat-exchanger temperature of $T_a$ = 300K in all cases. The acoustic perturbation is initialized with a standing wave of 0.5 kPa of amplitude and the source terms 
(\ref{eq:sourceTerms}) are applied from the beginning. The former is only used to reduce the duration of the transient and is not required to achieve acoustic energy growth (see section \ref{Results::EngineStartup}). Once a limit cycle is reached, two successive grid refinement steps are carried out, resulting in grid B and C. At each step the resolution was increased in the axial direction, especially around the sharp edges, and systematically doubled in the azimuthal direction. The mesh size in the radial direction is then adjusted and/or increased to optimize the cells aspect ratio. The sensitivity of the wave-induced Reynolds stresses to these changes (shown later) is used as a metric for grid convergence and only carried for calculations at $T_h$=500 K (table \ref{tbl:mesh_temperature_settings}).

% will be shown. The presence of shear-layer instabilities in the fully nonlinear regime (not shown in the present paper) and the near-transitional Stokes Reynolds numbers in the feedback inertance loop and in the neck of the resonator, suggest more constraining grid resolution requirements which will be explored in future work.

% % Numerics on that given mesh, code description, time advancement and spatial discretization, MPI stuff
The governing equations for mass, momentum and total energy are solved in the finite-volume unstructured code \emph{CharLES$^X$}
developed as a joint-effort project among researchers at Stanford University. The flux reconstruction method is grid-adaptive
at the preprocessing stage and solution-adaptive at run-time. It blends a high-order polynomial interpolation scheme (up to
fourth-order on uniform meshes) with a lower-order scheme to ensure numerical stability in areas of low grid quality 
\citep{HamMIM_2007_bookchpt}. A second order ENO reconstruction is adopted within the thermal buffer tube to control unwanted
oscillations in the solution caused by the application of the source terms \eqref{eq:sourceTerms}. The discretized system of 
equations is integrated in time with a fully-explicit, third-order Runge-Kutta scheme. The code is parallelized using the Message
Passing Interface (MPI) protocol and highly scalable on a large number of processors. The adoption of computationally intensive 
discretizations such as ENO in a limited portion of the domain has lead to a load-balancing problem that required a
volume-based dual-constrained partitioning \citep{KarypisK_ProcSComp_1998,SchloegelKK_ProcEuroPar_1999} to recover 
acceptable performance.

\begin{table}
  \centering
  \caption{Parameter space for numerical simulations carried out to limit cycle. Total 
  number of control volumes $N\textrm{cv}$, control volumes in the azimuthal direction $N_\theta$, temperature in the HHX $T_h$. For all cases $T_a = 300$K. Three meshes with increasing level of resolution and quality (from Grid A to C) are considered. Computations performed (x) and not performed ($\,\cdot\,$).}
  \label{tbl:mesh_temperature_settings}
  \begin{tabular}{cccccc}
  \\
     Grid Type  & $N_\textrm{cv}$ & $N_\theta$ & $T_h = 460$K &  $T_h = 480$K &  $T_h = 500$K  \\
       \hline
      A & 0.46m  & 20 &  (x)   &  (x)  &  (x) \\
      B & 1.25m  & 40 &  ($\,\cdot\,$)  &  ($\,\cdot\,$)  &  (x) \\
      C & 5.08m  & 80 &  (x)  &  ($\,\cdot\,$)  &  (x) \\ 
  \end{tabular}
\end{table}

% \section{Results}

% \caption{Time series of gauge pressure (a) and velocity (b) taken along the line of symmetry of the resonator at $x = -0.92 $ m (\textendash$\,$\textendash) and $x = 0.0425$ (\textemdash). Positive velocity corresponds to fluid moving towards the REG/HX unit, gauge pressure is reported in percentage with respect to atmospheric.}consistent with the porosity and wire size of the regenerator and heat-exchanger.

\section{Engine Start-Up} \label{Results::EngineStartup}

\begin{figure}
 \centering
 \includegraphics[keepaspectratio=true,width=0.95\linewidth]{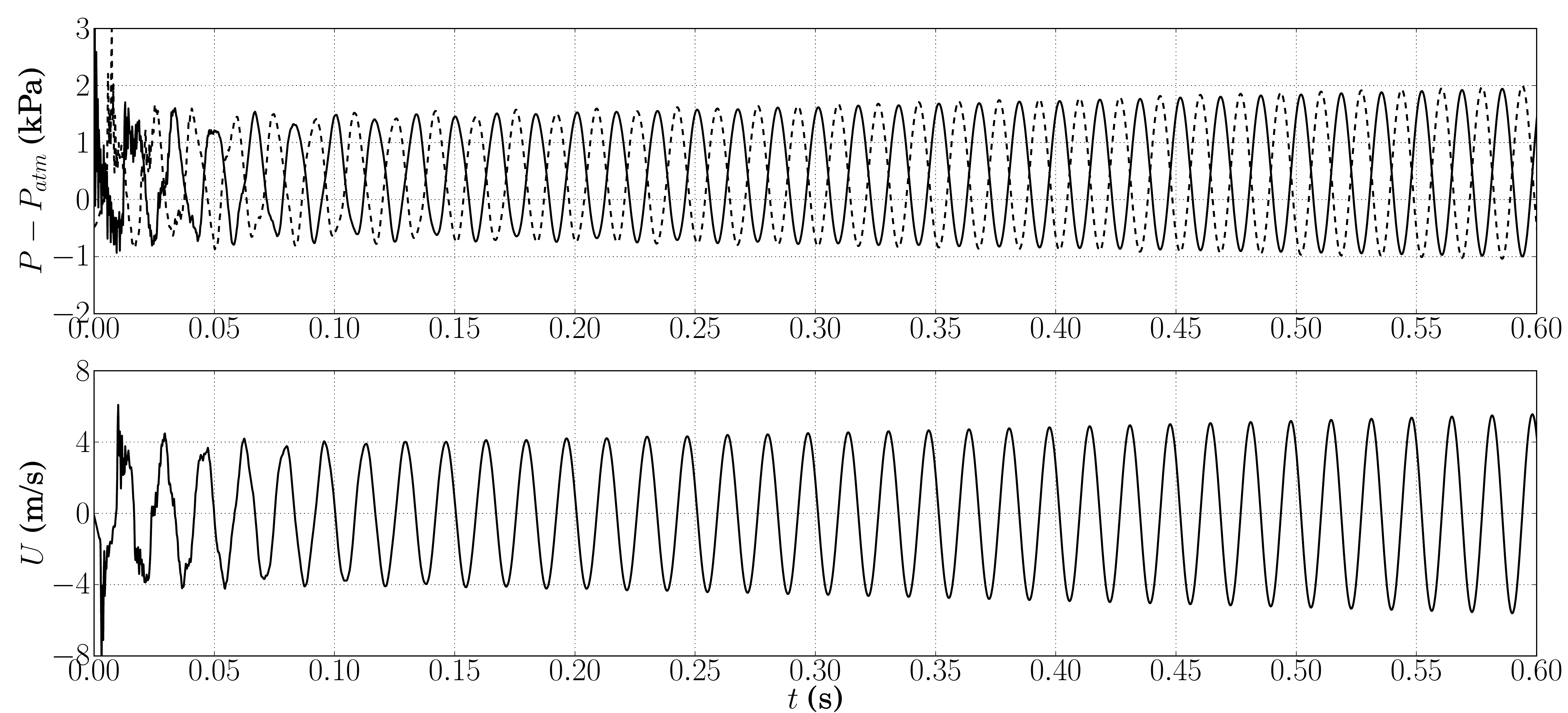}
 \put(-370,90){(a)}
 \put(-370,10){(b)}
\caption{Time series of pressure (a) along the centerline, $r=0$, at $x = -1.8825 $ m (\textendash$\,$\textendash) and $x = 0.0425 $m (\textemdash), and axial velocity (b) at $x = -0.92 $ m (\textemdash) for case $T_h = 500$ K on grid A.}
\label{fig:pressureVelTimeseries}
\end{figure}

\begin{figure}
\centering
\includegraphics[keepaspectratio=true,width=0.95\linewidth]{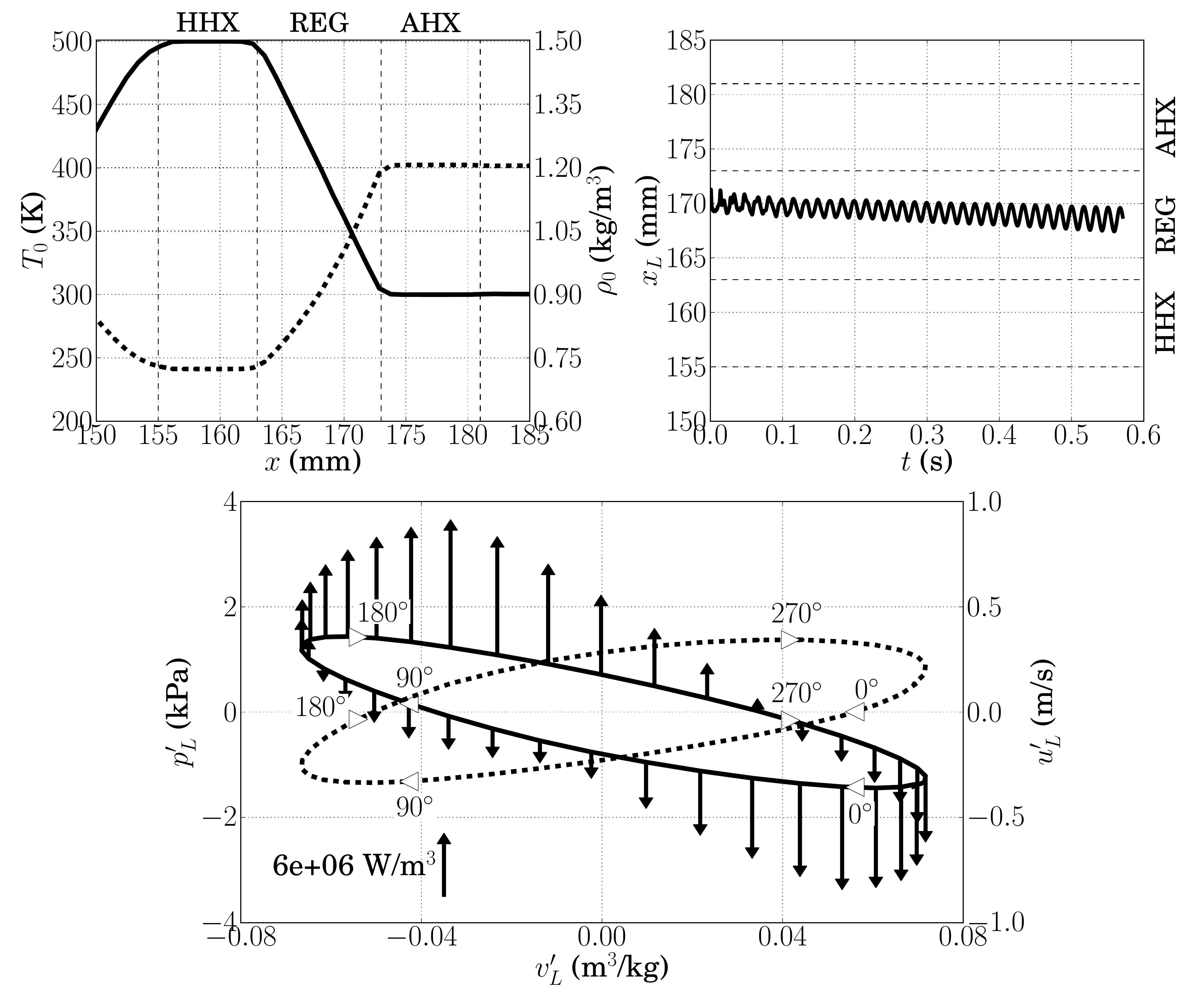}
\put(-365,170){(a)}
\put(-5,170){(b)}
\put(-50,20){(c)}
\caption{Base temperature (\textemdash) and density profiles (-$\,$-) taken along the centerline in the REG/HX
unit at $t=0.55$s for $T_h=500$ K, grid A (see table \ref{tbl:mesh_temperature_settings}) (a); Lagrangian fluid 
parcel axial position $x_L$ starting at $x=0.171$~m at $t = 0$ (b); fluctuating pressure, $p'_L$, (\textemdash) 
and Lagrangian velocity, $u'_L$,  (-$\,$-) plotted against specific volume fluctuation, $v'_L$, over one acoustic cycle $t=0.55$s
(c) with heat release fluctuation, ${S_h}'$ \eqref{eq:energySource_heat} shown with arrows (oriented upwards for heat
addition $S'_h>0$ and downwards for removal $S'_h<0$ with scale in figure). Both cycles are traversed clockwise as shown by the
phase markers.}
\label{fig:tempDensity_parcelTracking}
\end{figure}

%% Initial conditions
In all of the numerical trials performed, the abrupt activation of the source terms \eqref{eq:sourceTerms} alone in a 
quiescent flow provides a sufficiently intense initial disturbance ($\sim$1kPa) to trigger the thermoacoustic 
instability, leading to the production of acoustic energy in the system. Several attempts have been made to
reduce the initial pressure amplitude but have not been successful. In order to rapidly damp the broadband 
component of such disturbance (figure~\ref{fig:pressureVelTimeseries}) all cases are initiated with 
a half-wavelength pressure distribution of 0.5kPa in amplitude, with base pressure and 
temperature of 101,325 Pa and 300 K. Pressure amplitude initially grows exponentially for all locations at the same rate
and at the same frequency (at $\sim$60Hz). A standing wave develops in the resonator with growing amplitude and increasing base pressure level (DC pressure mode). The latter is caused by 
the expansion of the fluid in the TBT in contact with the HHX (see figure~\ref{fig:tempDensity_parcelTracking}a 
for x $<$ 0.155 m). Both base and acoustic pressure amplitude settle at a constant value as the second ambient heat 
exchanger picks up the excess heat. These are nonlinear effects that will be discussed later in 
section~\ref{Results::NonlinearRegime}. In the following we restrict the analysis of the generation and propagation of acoustic energy in the system during the start-up phase to linear acoustics.

\subsection{Thermodynamic Cycle} \label{sec:thermodynamic_cycle}

The driver of the thermoacoustic instability, converting heat into acoustic power, is the mean temperature 
gradient imposed in the REG/HX unit (figure~\ref{fig:tempDensity_parcelTracking}a). Insight into the fundamental 
energy conversion mechanisms can be gained by looking at the evolution of a Lagrangian fluid parcel in the 
regenerator interacting with the acoustic field. The slight drift 
in the direction of the mean temperature gradient (figure~\ref{fig:tempDensity_parcelTracking}b) is a nonlinear effect known as acoustic streaming (discussed 
later in section~\ref{Results::NonlinearRegime}), which can be ignored at this stage.

% there is a thermodynamic cycle
The fluid parcel in the regenerator experiences a thermodynamic cycle converting heat into acoustic power, which is neither the 
ideal Stirling or Carnot cycle (figure~\ref{fig:tempDensity_parcelTracking}c). For example,
purely isochoric transformations, present in the ideal Stirling cycle, are not possible due to the sinusoidal waveform
of the acoustic velocity and pressure. Moreover, isentropic transformations, present in the ideal Carnot cycle, are not 
possible due to the heat-exchange and viscous losses in the REG/HX unit. However, analogously to the ideal Stirling cycle, 
most of the heating and cooling occurs, respectively, during the expansion and compression stages. The heat transfer 
model in \eqref{eq:energySource_heat} assumes perfect thermal contact and is therefore likely that its adoption leads 
to an overestimation of the real thermoacoustic response under comparable conditions.

% A qualitative explanation for the acoustic energy production can be provided building upon the ideas in \cite{Ceperley_1979_JAcoustSA} that led to the design of the traveling-wave thermoacoustic engine.
Due to the orientation of the background temperature gradient ($d\,T_0(x)/dx<0$, figure~\ref{fig:tempDensity_parcelTracking}a), a fluid parcel in the regenerator that is displaced towards the hot heat exchanger ($u'<0$) will experience heating. This occurs with a given phase-lag with respect to the velocity depending, in particular, on the nature of the heat-transfer. In our case the heat addition, $S_h'>0$, peaks $\sim 90^\circ$ after the maximum negative velocity (when the parcel comes to rest) and so does the positive pressure fluctuation, $p'>0$ (figure~\ref{fig:tempDensity_parcelTracking}c). The opposite occurs when the fluid is displaced towards the ambient heat exchanger. Consistently with the energetic considerations underlying the Rayleigh criterion, $\overline{S_h'\,p'}>0$, the observed phase differences suggest the presence of positive acoustic energy production. The phasing between velocity and pressure fluctuations is consistent with a standing wave, short of approximately 15$^{\circ{}}$. This slight phase difference contributes to a negative 
correlation between $u'$ and $p'$, i.e. the left-traveling wave propagating through the REG/HX is more intense than the right-traveling wave. Acoustic power is therefore being produced.

\begin{figure}
\centering

\includegraphics[width=.7\linewidth]{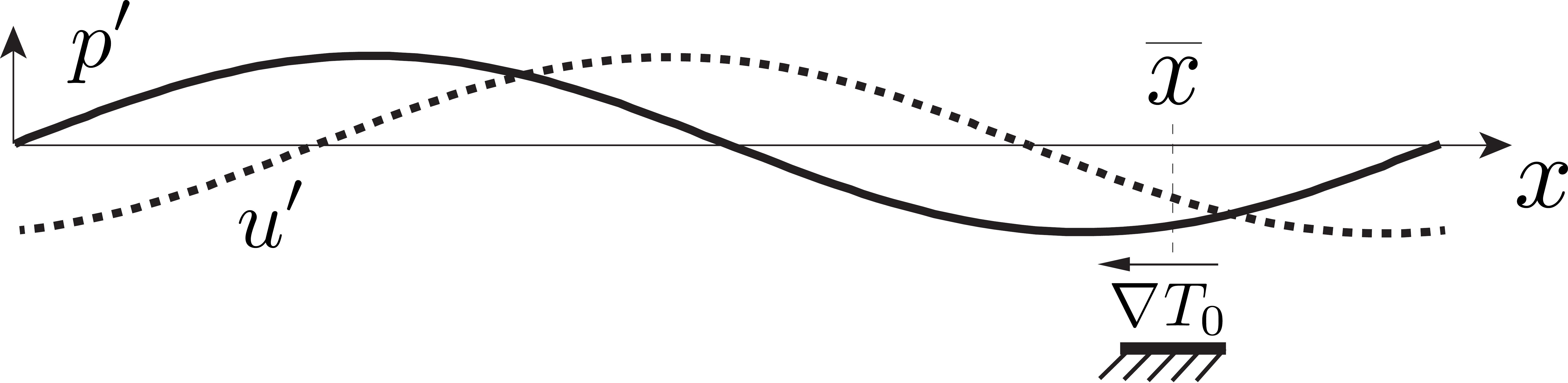}
\vspace*{-0.12cm}

\includegraphics[width=0.95\linewidth]{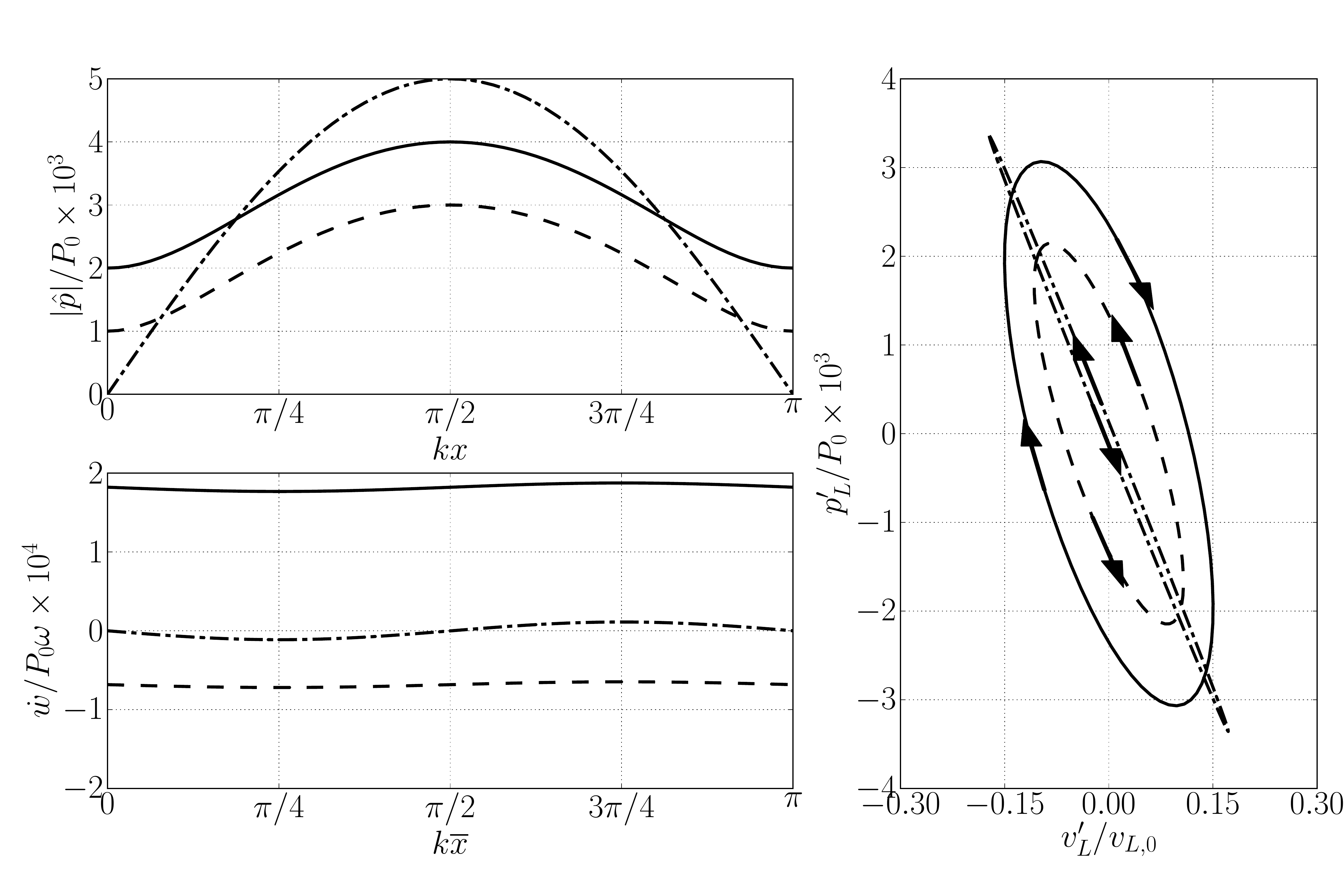}
\put(-365,135){(a)}
\put(-365,25){(b)}
\put(-2,40){(c)}
\vspace*{-0.5cm}
\caption{Results from the linear Lagrangian model (section \ref{sec:thermodynamic_cycle}) of a one-dimensional theoretical thermoacoustic engine composed of a generic
plane wave interacting with a negative temperature gradient located at $\overline{x}$ (top). Imposed pressure distribution as a function of $k\,x$ (a), overall cycle-averaged work (i.e. generated acoustic power) as a function of $k\overline{x}$ (temperature gradient location) (b), and Lagrangian thermodynamic cycle extracted in the temperature gradient region for $k\,\overline{x} = \pi/4$ (c). Results for plane waves of amplitude  $p^-  =  3\times10^{-3}\,P_0$, $p^+  =  1\times10^{-3}\,P_0$ (\textendash\textendash); $p^+ = p^- = 2.5 \times10^{-3}\,P_0$ (\textendash$\,\cdot\,$\textendash); $p^-  =  1\times10^{-3}\,P_0$, $p^+  =  3\times10^{-3}\,P_0$ (\textendash$\,$\textendash) for $P_0 = \rho_0\,a_0^2$. For all cases $\phi^- = \phi^+ = - \pi/2$.}
 \label{fig:swiftTheoreticalEngine}
\end{figure}

Inspired by \cite{Swift_1988_JAcoustSA}'s theoretical thermoacoustic engine (figure \ref{fig:swiftTheoreticalEngine}, top), a one-dimensional analytical model is derived in the following to rigorously explain the interaction between a mean temperature gradient and a Lagrangian fluid parcel in a generic planar acoustic wave field.
%\begin{align}
%p'  = p^- \textrm{cos}(kx+\omega\,t+\phi^-)+p^+ \textrm{cos}(kx-\omega\,t+\phi^+) \\
%u' = -u^- \textrm{cos}(kx+\omega\,t+\phi^-)+u^+ \textrm{cos}(kx-\omega\,t+\phi^+),
%\end{align}
The latter can expressed as the linear superposition of a left- and a right-traveling wave, which in complex form reads
\begin{align} \label{eq:leftandrighttravelingwave}
\hat{p}(x) = p^- e^{i\left[k\,x+\phi^-\right]} + p^+ e^{i\left[-k\,x-\phi^+\right]} \\
\rho_0 \; a_0 \;\hat{u}(x) =  - p^- e^{i\left[k\,x+\phi^-\right]} + p^+ e^{i\left[-k\,x-\phi^+\right]},
\end{align}
where $p^{+/-} $ is the amplitude of the right/left- traveling wave and $k$ is the wave number. The acoustic pressure and velocity in time are given based on the convention $p'_a = \hat{p}(x)\expiomega$ and $u'_a = \hat{u}(x)\expiomega$, where $i$ is the imaginary unit and the base impedance $\rho_0\,a_0$ is given by the state $\{\rho_0$, $T_0$, $P_0\}$.  

Let $x_p = \overline{x} + x_p'$ be the instantaneous position of a fluid parcel oscillating with small displacements $x_p'$ about the position $\overline{x}$, where a linear temperature gradient is locally imposed (figure \ref{fig:swiftTheoreticalEngine}). The fluid parcel velocity can be approximated, based on the assumption $k\,\textrm{max}(x_p') << 1$, as
\begin{equation}
\frac{d }{d t} x'_p = u'(\overline{x}+x'_p,t) = \mathcal{R}\{\hat{u}(\overline{x}+x'_p) e^{i\,\omega\,t} \} \simeq \mathcal{R}\{\hat{u}(\overline{x}) e^{i\,\omega\,t} \}
\end{equation}
yielding the relation 
\begin{equation} \label{eq:fluid_particle_vel}
\hat{x}_p = \hat{u}(\overline{x})/i\omega.
\end{equation}
Introducing a Lagrangian base state $\rho_{L,0},P_{L,0},T_{L,0}$ (specified later) for the fluid parcel density, entropy and temperature,
\begin{align}
\rho_L = \rho_{L,0} + \mathcal{R} \left\{\hat{\rho}_L\,\expiomega \right\} \\
s_L = s_{L,0} + \mathcal{R} \left\{\hat{s}_L\,\expiomega \right\} \\
T_L = T_{L,0} + \mathcal{R} \left\{ \hat{T}_L\,\expiomega \right\},
\end{align}
and substituting into the heat transfer equation, expressed in terms of entropy with heat source modeled based on \eqref{eq:energySource_heat} and evaluated at the parcel position $x_p$, yields
\begin{equation} \label{eq:basic_heat_transfer}
\rho_L T_L \frac{d s_L}{dt} = - \alpha_T \left[ T_L - T_0(x_p) \right]
\end{equation}
where the (total) time derivative on the l.h.s., applied to the Lagrangian fluid parcel's specific entropy, has replaced the material derivative of the Eulerian entropy field.

Substituting Gibbs' relationship linearized about the Lagrangian base state $T_{L,0},p_{L,0}$(=$p_0$),
\begin{equation} \label{eq:Gibbs_relation}
ds_L = \frac{c_p}{T_{L,0}}dT_L - \frac{R}{p_0}dp_L
\end{equation}
into \eqref{eq:basic_heat_transfer}, which is also linearized assuming $\rho_L T_L \simeq \rho_{L,0} T_{L,0}$, yields
\begin{equation} \label{eq:basic_heat_transfer2}
\rho_{L,0} c_p \frac{dT_L}{dt} - \frac{d\,p_L}{dt} = - \alpha_T \left[ T_L - T_0(x_p) \right].
%\rho_{L,0} T_{L,0} \frac{d s_L}{dt} = - \alpha \left[ T_L - T_0(x_p) \right].
\end{equation}
The imposed mean temperature at the particle location, $T_0(x_p)$, can be expressed via the Taylor expansion 
\begin{equation} \label{eq:wall_temp}
T_0(x_p) = T_0(\overline{x})+ x_p' \nabla\, T_0,
\end{equation}
which is exact in the case of linear temperature profile. Substituting \eqref{eq:wall_temp} into \eqref{eq:basic_heat_transfer2} yields
\begin{equation} \label{eq:basic_heat_transfer3}
\rho_{L,0} c_p \frac{dT_L}{dt} - \frac{d\,p_L}{dt} = - \alpha_T \left[ T_L - T_0(\overline{x})- x_p' \nabla\, T_0 \right]
\end{equation}

Defining the Lagrangian based state for temperature such that $T_{L,0} = T_0(\overline{x})$, assuming $p_L = p$ and switching to complex form
\begin{equation} \label{eq:basic_heat_transfer4}
\rho_{L,0} c_p i\omega \hat{T}_L - i\omega \hat{p} = - \alpha_T \left[ \hat{T}_L - \hat{x}_p \nabla\, T_0 \right]
\end{equation}
where the Lagrangian base density is $\rho_{L,0} = p_0 / (R\,T_{L,0})$. 

If the acoustic field is assigned, so are the complex pressure amplitude $\hat{p}$ and particle displacement $\hat{x}_p= \left(i\omega\right)^{-1}\hat{u}(\overline{x})$, which then allows to solve for $\hat{T}_L$ from \eqref{eq:basic_heat_transfer4}. The density of the Lagrangian parcel can then be calculated from the linearized equation of state
\begin{equation} \label{eq:LagrDensity}
\hat{\rho}_L = \frac{\hat{p} - \rho_{L,0} R \hat{T}_{L}}{R\,T_{L,0}}.
\end{equation}

The work done by the fluid parcel on the surrounding ambient per unit time (generated acoustic power) is
\begin{equation}
 \dot{w} = - \overline{\frac{p}{\rho} \frac{D \rho}{D t}} \simeq - \frac{1}{\rho_{L,0}} \frac{1}{2} \mathcal{R} \{ \hat{p} \left(i\,\omega \hat{\rho}_L \right)^* \}.
\end{equation}
Depending on the nature of acoustic wave being imposed (figure \ref{fig:swiftTheoreticalEngine}a), ranging from purely left-traveling to purely right-traveling, a different phasing between $\hat{p}$ and $\hat{u}$ is achieved, determining $\dot{w}$ (figure \ref{fig:swiftTheoreticalEngine}b). In the case of a standing wave, acoustic energy will be absorbed ($\dot{w}<0$) or generated ($\dot{w}>0$) depending on the location of the temperature gradient, $\overline{x}$. For a sufficiently high amplitude of the right-, $p^+>>p^-$, (or left-, $p^->>p^+$) traveling wave, acoustic power is ultimately only absorbed (or generated) for any $\overline{x}$ (figure \ref{fig:swiftTheoreticalEngine}b). This shows that an acoustic wave traveling in the same or opposite direction of an imposed mean temperature gradient (applied over a region small compared to the wavelength) will be amplified or absorbed. Moreover, the acoustic power associated with the energy conversion occurring in an (almost) purely traveling wave is remarkably 
higher (see area enclosed by the $p-v$ diagrams in figure \ref{fig:swiftTheoreticalEngine}c) than the one of a standing wave of comparable amplitude. This confirms \cite{Ceperley_1979_JAcoustSA}'s seminal intuition that led to the revolutionary concept of traveling-wave energy conversion, trumping thereafter designs based on standing waves.

% Works by Sujith and Juniper have argued that non-modal growth mechanisms can be found in TAE. In our numerical experiment, the linearity of the low-order model used for the heat-transfer determines . nonlinear effects such as vortex-shedding from a heated grid or out of the porous media are not taken into account and might lead to different dynamics for thermoacoustic instability.

\begin{figure}
\centering
\includegraphics[keepaspectratio=true,width=.95\linewidth]{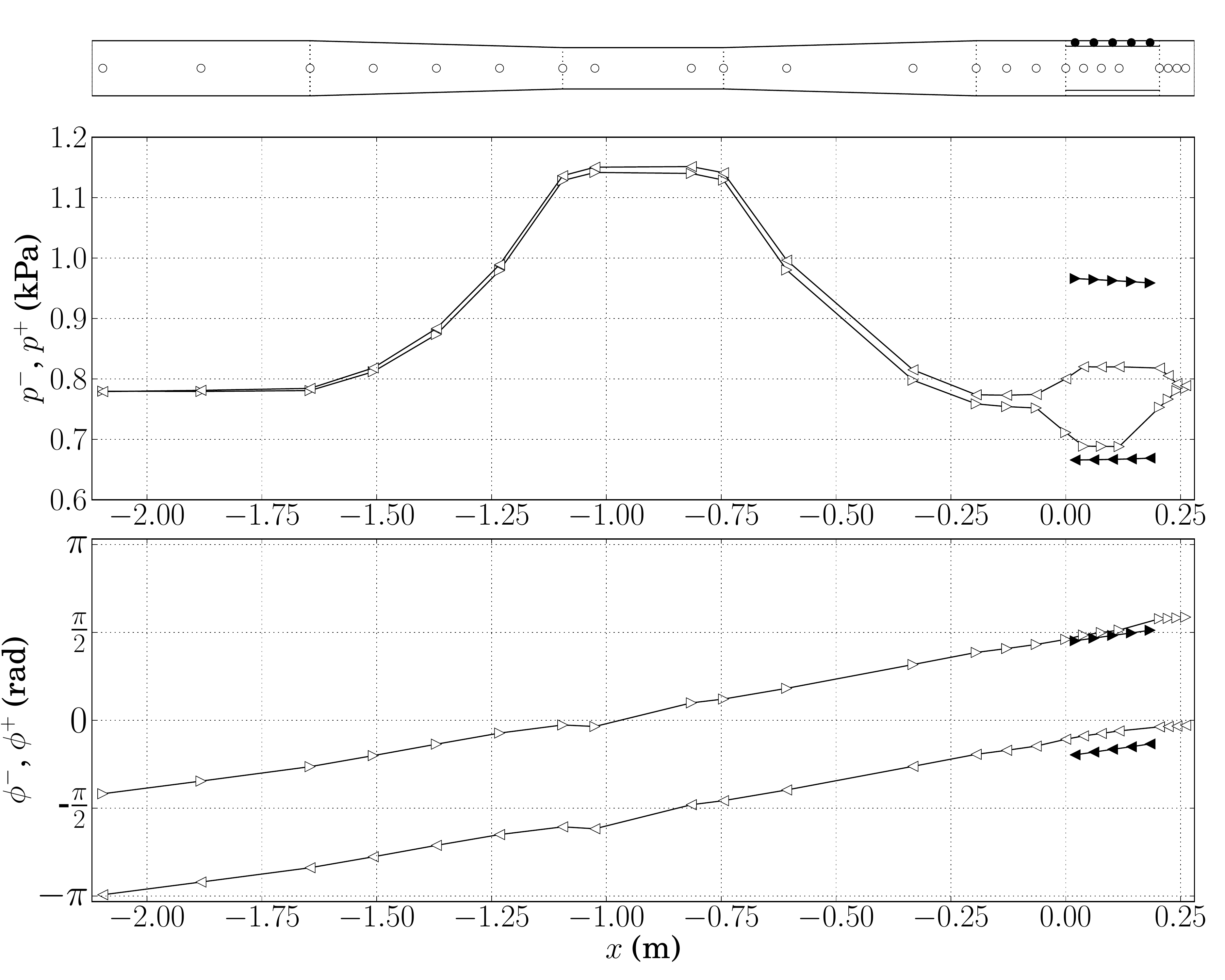}
\put(-367,145){(a)}
\put(-367,20){(b)}
\caption{Left ($-$, $\triangleleft$, $\blacktriangleleft$) and right ($+$, $\triangleright$, $\blacktriangleright$) traveling-wave amplitudes,  $p^\pm$, (a) and phases,  $\phi^\pm$, (b) calculated based on the linear approximation \eqref{eq:p_and_u_prime_LeftRightTraveling} for time $t = 0.55$ s for $T_h = 500$ K case on grid A. The locations where data is extracted are shown in the top figure. White and black symbols correspond to values extracted on the centerline and in the feedback inertance, respectively.}%   \caption{Left  (\textemdash$\,$\textemdash)
\label{fig:acousticNetwork}
\end{figure}
% , %{\bf I think that linear theory works because the heat-transfer model is linear. If the heat-transfer was resolved I think there might be nonlinear effects such as vortex-shedding. Can we use this consideration to bring up the papers by  that argued for non-modal growth?}

\subsection{Acoustic Network of Traveling Waves} \label{Results::LeftRight_TravelingWavesAnalysis}

In spite of the finite amplitude of the initial perturbation (exceeding 1\% of the base pressure) and the immediate establishment of nonlinear effects, the exponentially growing acoustic amplitude (with uniform growth rate in the entire system) suggests that the system-wide behavior in the start-up phase can be analyzed by invoking linear acoustics. The low frequencies observed in the numerical simulations and the high-aspect ratio of the resonator (with lowest cut-on frequency $\sim$1.7 kHz) allows us to neglect radial or azimuthal acoustic modes at the resonator scale. This suggests that, as a first approximation, our analysis can be restricted to planar waves. The nature of the thermoacoustic instability (section \ref{sec:thermodynamic_cycle}) suggest that the REG/HX unit acts as an amplifier of left-traveling waves, which propagate into the resonator and are reflected back; upon returning to the REG/HX unit as right-traveling waves, they are expected to propagate both through the pulse tube and in the feedback 
inertance (figure \ref{fig:computationalsetupNijeholt}), being absorbed in the former case (figure \ref{fig:swiftTheoreticalEngine}c) and propagating freely in the latter. The acoustic power propagating through the inertance is looped back into the REG/HX unit via the compliance, hence creating a network of self-amplified traveling-waves.

This picture can be confirmed with the aid of the instantaneous numerical data. An exact local decomposition in terms of left $(-)$ and right $(+)$ traveling waves
\begin{subnumcases}{\label{eq:p_and_u_prime_LeftRightTraveling}}
  p'(t) = p^{-}\;f(\omega\;t+\phi^{-})+p^{+}\;f(-\omega\;t+\phi^{+}), \\
  u'(t) = -u^{-}\;f(\omega\;t+\phi^{-})+u^{+}\;f(-\omega\;t+\phi^{+}),
\end{subnumcases}
% As previously stated, during the linear regime, the acoustic perturbation is a pure tone at 59.8 Hz growing exponentially over time scales longer than the acoustic period.  (with good approximation away from walls) 
can allow for the direct evaluation of the amplitudes $p^{\pm}$ and $u^{\pm}$ and phases $\phi^\pm$ of purely traveling waves of a given generic waveform $f()$ (periodic function of period $2\,\pi$). In our case, the angular frequency $\omega$ is much larger than the growth rate (discussed in section \ref{Results::linearModel}), allowing to ignore the variations of pressure and velocity amplitudes over one acoustic period as well as the change in base impedance, $\rho_0\,a_0$, due to a DC mode in pressure (discussed above). The following analysis is restricted to the start-up phase and for locations in the engine outside of the REG/HX unit where the strong mean temperature causes non-negligible spatial gradients of the base impedance and the isentropic wave propagation assumption to be violated.

\begin{figure}
  \centering
  \includegraphics[keepaspectratio=true,width=\linewidth]{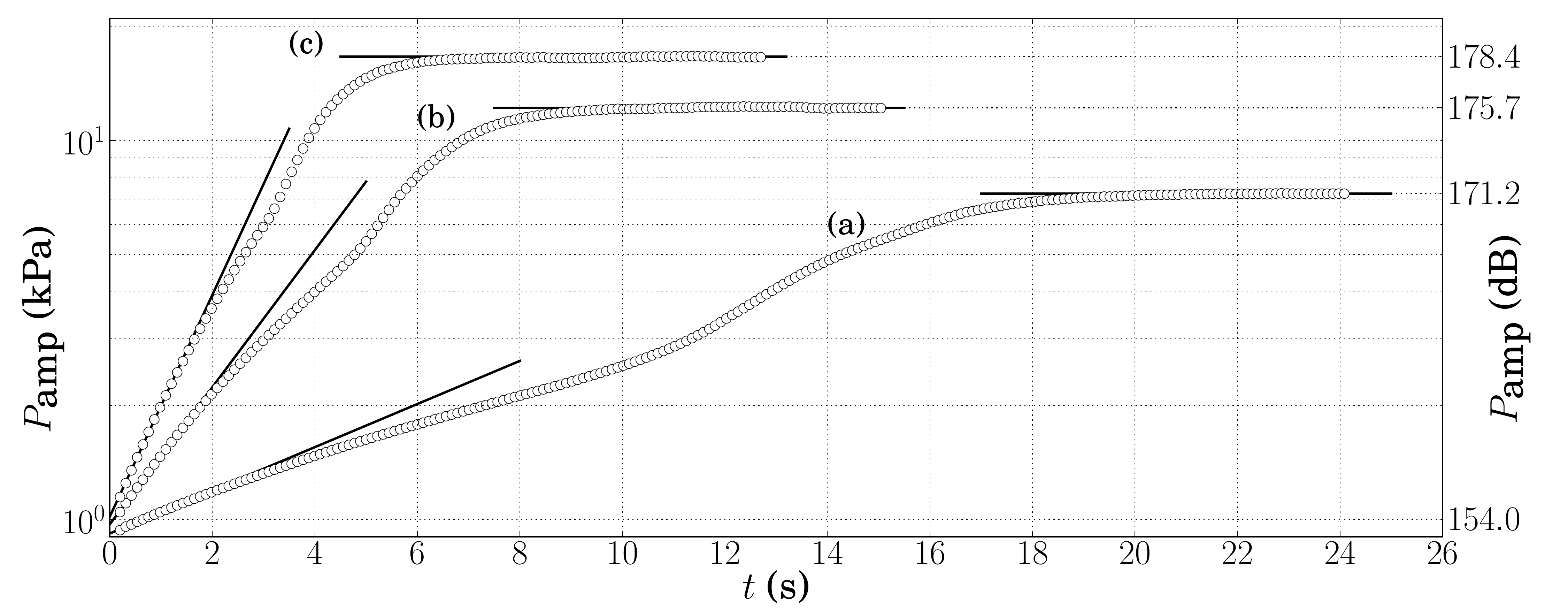}
  \caption{Time series of pressure amplitude (kPa) on the left end of the resonator and corresponding sound pressure level (+dB) for $T_h = 460$K (a), $T_h = 480$K (b) and $T_h = 500$K (c) on grid A (see table \ref{tbl:mesh_temperature_settings}). Decaying case for $T_h=440$K not shown.}
  \label{fig:TimesSeriesSemiLog}
\end{figure}

The acoustic perturbation \eqref{eq:p_and_u_prime_LeftRightTraveling} can be rewritten in the form of a complex Fourier series
\begin{subnumcases}{\label{eq:p_and_u_prime_Fourier}}
\DoubleInfSum \hat{p}_k\;e^{i\,k\,\omega\;t} = p^{-}\,\DoubleInfSum \hat{f}_k \;e^{i\,k\left[\omega\;t+\phi^{-}\right]}+p^{+}\,\DoubleInfSum \hat{f}_k^* e^{i\,k\left[\omega\;t-\phi^{+}\right]}, \\
\rho_o\; a_o \DoubleInfSum \hat{u}_k\;e^{i\,k\,\omega\;t} = -p^{-}\,\DoubleInfSum \hat{f}_k \;e^{i\,k\left[\omega\;t+\phi^{-}\right]}+ p^{+}\,\DoubleInfSum \hat{f}_k^* e^{i\,k\left[\omega\;t-\phi^{+}\right]}.
\end{subnumcases}
where $p^{\pm} = \pm \rho_0\,a_0 u^{\pm}$ and the superscript $^*$ indicates the complex conjugate. Letting the $m$-th mode be any non-zero Fourier component, the unknowns $p^{\pm}$ and $\phi^{\pm}$ are easily determined by isolating such mode from \eqref{eq:p_and_u_prime_Fourier}, yielding
\begin{subnumcases}{\label{eq:p_and_u_prime_FourierOneComponent}}
  \hat{p}_m =  \hat{f}_{m} \left[ p^{-}\,e^{i\,m\,\phi^{-}} \right] \; + \hat{f}_m^* \left[ p^{+} e^{-i\,m\,\phi^{+}} \right], \\
  \rho_o\,a_o \hat{u}_m = -\hat{f}_m \left[ p^{-} \;e^{i\,m\,\phi^{-}} \right] + \hat{f}_m^* \left[ p^{+} e^{-i\,m\,\phi^{+}}\right].
  \end{subnumcases}
The amplitudes and phases of the modes in \eqref{eq:p_and_u_prime_LeftRightTraveling} can be obtained from the magnitude and phases of the complex unknowns grouped into squared brackets above, yielding
\begin{subnumcases}{\label{eq:p_and_u_prime_FourierSolution}}
   p^{-} = |\hat{p}_m-\rho_0\,a_0\hat{u}_m| / | 2\hat{f}_m | \\
   p^{+} = |\rho_0 a_0 \hat{u}_m + \hat{p}_m| / | 2\hat{f}_m^* |.
 \end{subnumcases}
The phases $\phi^{-}$ and $\phi^{+}$ are then readily extracted from \eqref{eq:p_and_u_prime_FourierOneComponent} given the pressure amplitudes $p^{-}$ and $p^{+}$.

This procedure is applied to the discrete set of points in figure \ref{fig:acousticNetwork}(top) located along the resonator axis and around the pulse tube. Results show that left-traveling waves leaving the REG/HX unit propagate into the resonator and are reflected back with slightly lower amplitude due to losses in the resonator (figure \ref{fig:acousticNetwork}a). Consistently with the one-dimensional approximation in \eqref{eq:p_and_u_prime_LeftRightTraveling} the acoustic power can be expressed as
\begin{equation} \label{eq:acoustic_power}
 \dot{E}^+_{a} = \int_A \overline{p'\;u'}\,dA = \frac{A}{\rho_0\;a_0} \left[ {p^{+}}^2-{p^{-}}^2 \right] \DoubleInfSum {\hat{f}_k}^* \hat{f}_k.
\end{equation} % Pa m^3/s = N * m/s = W
providing an energetic interpretation to the imbalance $p^{-} \gtrless p^{+}$. The data (figure \ref{fig:acousticNetwork}a,b) confirms that, as anticipated earlier in this section, the acoustic power generated in the REG/HX is fed back to it the after being channeled through the inertance, where ${p^{+}}^2\,>>\,{p^{-}}^2$ (directly responsible for nonlinearities such as Gedeon streaming, analyzed later in section \ref{Results::NonlinearRegime}). The spatial distribution of $\phi^{-}$ and $\phi^{+}$ (figure \ref{fig:acousticNetwork}b) is consistent with a standing wave at the resonator scale with a slight deviation in the REG/HX and in the inertance. Such difference is sufficiently small to suggest that it is in fact the same planar wave propagating through the REG/HX and the inertance in each direction at once but with part of its acoustic power amplified or damped in the former depending on the direction of propagation. Overall, the present results confirm the qualitative picture outlined earlier in this section, explaining the instability mechanisms leading to the acoustic energy growth (figure \ref{fig:TimesSeriesSemiLog}).

\subsection{System-wide Linear Modeling}  \label{Results::linearModel}

Results shown so far suggest that nonlinearities do not play an important role in explaining the acoustic energy propagation and amplification mechanisms during the start-up phase. However, given the high amplitude of the initial perturbation ($\sim 1$kPa) and the presence of complex geometrical features such as the sharp edges of the pulse tube (inducing vortex shedding from the start), it is important to assess to what extent a system-wide linear model is able to quantitatively explain the observed instability.

\begin{figure}
  \centering
  \includegraphics[keepaspectratio=true,width=0.95\linewidth]{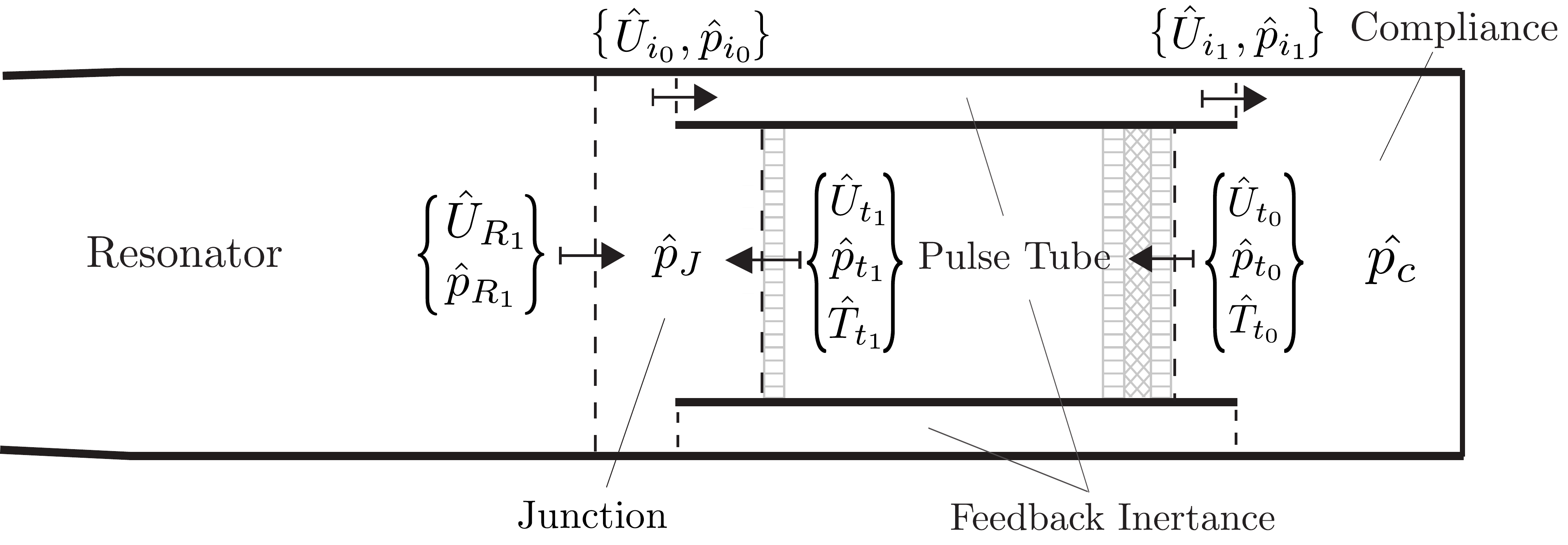}
%   \includesvg{./figures/lumped-parameter-modeling/Sequential_RottIntegration}
  \caption{Sketch illustrating the subdivision of the engine into control volumes representing different components (cfr figure \ref{fig:computationalsetupNijeholt}): compliance ($c$), pulse tube ($t$), feedback inertance ($i$), junction ($J$), and resonator ($R$). Control surfaces are shown with dashed lines marking the beginning ($0$) and the end ($1$) of one-dimensional flow segments (oriented according to arrows) modeling the pulse tube, inertance and resonator. Complex pressure, volumetric flow rate and temperature are indicated with $\hat{p}$, $\hat{U}$, and $\hat{T}$ respectively. References to the thermal buffer tube are dropped in the context of the analysis of the start-up phase since nonlinear effects can be neglected. The left end of the control volume representing the junction is located at $x$ = -23 mm.}
\label{fig:linearStabilityAnalysis}
\end{figure}

Building upon well-established linear modeling approaches \citep{Rott_ZAMP_1969,WardS_JAcouSocAm_1994,deWaele_2009_JSV}, the engine is divided into a collection of control volumes (figure \ref{fig:linearStabilityAnalysis}), representing different components, each modeled as a one- or zero-dimensional element, exchanging acoustic power and mass with adjacent components. Data from the Navier-Stokes simulations suggests that the pressure field is uniform within the compliance, consistent with \cite{deWaele_2009_JSV}'s modeling choices. By imposing the conservation of mass and assuming an isentropic relation between volume-averaged density and pressure variations one obtains, in the time domain, 
\begin{equation}{\label{eq:complianceODE_intime}} 
\frac{d}{dt} p'_{c}= w_{c_0} \Big[ U'_{i_1} - U'_{t_0} \Big]
\end{equation}
where $p'_{c}$, $U'_{i_1}$ and $U'_{t_0}$ are, respectively, the instantaneous fluctuating pressure in the compliance and flow rates exchanged with the inertance through surface $i_1$, and with the pulse tube through $t_0$ (figure \ref{fig:linearStabilityAnalysis}), and $w_{c_0} = \gamma P_0 / V_{c_0}$ where $P_0$ is the base pressure and $V_{c_0}$ the volume of the compliance.

The very small variation among the growth rates and frequencies extracted from the numerical simulations at the locations in figure \ref{fig:acousticNetwork} (not shown) suggests that normal modes can be assumed for all fluctuating quantities. By adopting the $e^{+i\sigma\,t}$ convention with $\sigma=-i\alpha+\omega$, where $\alpha$ and $\omega$ are, respectively, the growth rate and the angular frequency, \eqref{eq:complianceODE_intime} becomes
\begin{equation}{\label{eq:complianceODE_infreq}}
i\sigma\hat{p}_c = w_{c_0} \left[ \hat{U}_{i_1} - \hat{U}_{t_0} \right].
 \end{equation}
The same modeling approach is adopted for convenience at the junction where the conservation of mass reads
\begin{equation}{\label{eq:junctionODE_infreq}}
i\sigma\,\hat{p}_J = w_{J_0} \left[ \hat{U}_{R_1} - \hat{U}_{i_0} +  \hat{U}_{t_1} \right] \\
\end{equation}
with $w_{J_0} = \gamma P_0 / V_{J_0}$, where $V_{J_0}$ is the volume of the control volume modeling the junction (figure \ref{fig:linearStabilityAnalysis}).

Phase variations along the axial coordinate, $x$, are significant for the other components of the engine and the direct application of the complete set of linearized Euler equations is necessary. In all cases, the fluctuating field and the base state, defined by $\rho_0$, $T_0$, and $P_0$, are assumed to be exclusively a function of $x$.

For the resonator ($R$) and feedback inertance ($i$) isentropic wave propagation is assumed, yielding a simplified set of linearized equations for mass and momentum,
\begin{subnumcases}{\label{eq:linearEuler_isentropic}}
i\sigma \hat{p} = - \frac{\rho_0\,a_0^2}{A(x)} \frac{d\hat{U}}{dx} \\
i\sigma \hat{U} = - \frac{A(x)}{\rho_0} \frac{d\hat{p}}{dx}
\end{subnumcases} 
valid for a variable cross-sectional area distribution $A(x)$, where $a_0 = \sqrt{\gamma\,R\,T_0}$ is the speed of sound based on the base temperature. By introducing a spatial discretization, the set of equations \eqref{eq:linearEuler_isentropic} can be recast in algebraic form, as
\begin{equation} \label{eq:subset_of_diagonalized_systems_R}
\left( i\sigma \textbf{I} - \textbf{B}_{R} \right) \cdot\textbf{u}_R = 0 \\
\end{equation}
for the resonator, and
\begin{equation} \label{eq:subset_of_diagonalized_systems_i}
\left( i\sigma \textbf{I} - \textbf{B}_{i} \right) \cdot\textbf{u}_i = 0
\end{equation}
for the inertance, where $\textbf{I}$ is the identity matrix, $\textbf{B}$ is an operator discretizing the r.h.s. of \eqref{eq:linearEuler_isentropic}, and $\textbf{u}$ is a discrete collection of complex amplitudes for pressure and flow rate, specifically, $\textbf{u}_R = \{ \hat{\textbf{p}}_R,\hat{\textbf{U}}_R \}$ for the resonator, and $\textbf{u}_i = \{ \hat{\textbf{p}}_i,\hat{\textbf{U}}_i \}$ for the feedback inertance. 

The systems of equations \eqref{eq:subset_of_diagonalized_systems_R} and \eqref{eq:subset_of_diagonalized_systems_i} are isolated component eigenvalue problems (with boundary conditions to be specified) and their resolution, in the context of a system-wide linear stability analysis, is only meaningful if coupled with all of the other components in the engine, as discussed in the following.

The heat-transfer and drag in the pulse tube, and the presence of gradients of base density and temperature in the REG/HX unit, require variations of entropy to be explicitly accounted for. Replacing the conservation equation for the total energy with the transport equation for entropy, expressed in terms of temperature and pressure using Gibbs' relation, yields
\begin{subnumcases}{\label{eq:linearEuler_nonisentropic}}
i\sigma A\hat{\rho} + \frac{d \hat{U}}{d x} \rho_0 + \hat{U}\frac{d \rho_0}{d x} = 0 \\
i\sigma\hat{U} + \frac{A}{\rho_0} \frac{d\hat{p}}{dx}  =  - \frac{R_C}{\rho_0} \hat{U} \\
\rho_0\,C_p\left[i\sigma\,\hat{T} + \frac{\hat{U}}{A} \frac{dT_0}{dx}\right] - i\sigma\hat{p} = - \alpha_T \hat{T}
\end{subnumcases}% -\nu \frac{1}{r}\frac{\partial}{\partial r} \left( r \frac{\partial \hat{u}}{\partial r}\right)  
where the source term on the r.h.s of (\ref{eq:linearEuler_nonisentropic}$b$) is obtained by linearizing the drag-model (\ref{eq:sourceTerms}$a$), due to \cite{Organ_CUPress_1992} and the heat-transfer model on the r.h.s of (\ref{eq:linearEuler_nonisentropic}c) is the same one used in \eqref{eq:energySource_heat} \citep{BejanA_2004_HeatTransfer}. Such terms are only activated in the heat-exchangers and the regenerator. The spatial distribution of the base state in the pulse tube is taken from the numerical data at $t=0.55$ s (figure \ref{fig:tempDensity_parcelTracking}a), at the early stages of the start-up phase (figure \ref{fig:TimesSeriesSemiLog}).

Recasting the system of equations in \eqref{eq:linearEuler_nonisentropic} in diagonalized form yields
\begin{subnumcases}{\label{eq:linearEuler_nonisentropic_diagonalized}}
i\sigma\,\hat{p}  = \left[ \rho_0\,R\,\mathcal{B}_{\hat{T}\hat{U}} - \frac{P_0}{A} \frac{d}{dx} - \frac{R\,T_0}{A} \frac{d\rho_0}{dx} \right] \hat{U} + \left[\rho_0\,R\,\mathcal{B}_{\hat{T}\hat{T}}  \right] \hat{T}\\
i\sigma\,\hat{U} = \left[-\frac{A}{\rho_0} \frac{d}{dx} \right] \hat{p} + \left[-\frac{R_c}{\rho_0} \right] \hat{U} \\
i\sigma\,\hat{T} = \mathcal{B}_{\hat{T}\hat{U}}\,\hat{U} +  \mathcal{B}_{\hat{T}\hat{T}}\,\hat{T}
\end{subnumcases}
where
\begin{eqnarray}
&&\mathcal{B}_{\hat{T}\hat{U}} = -\frac{T_0}{A}\left[ \frac{1}{T_0}\frac{d\,T_0}{dx} + \left(\gamma-1\right)\frac{d}{dx} \right] \\ 
&&\mathcal{B}_{\hat{T}\hat{T}} =  - \frac{\gamma\,\alpha_T}{\rho_0\,C_p}
\end{eqnarray}
which, discretized in space, yields
\begin{equation}\label{eq:subset_of_diagonalized_systems_t}
\left( i\sigma \textbf{I} - \textbf{B}_{t} \right) \cdot\textbf{u}_t = 0
\end{equation}
where $\textbf{u}_t = \{\hat{\textbf{p}}_t, \hat{\textbf{U}}_t, \hat{\textbf{T}}_t\}$.  \\

The complete eigenvalue problem can finally be solved by coupling the isolated component eigenvalue problems \eqref{eq:complianceODE_infreq}, \eqref{eq:junctionODE_infreq}, \eqref{eq:subset_of_diagonalized_systems_R}, \eqref{eq:subset_of_diagonalized_systems_i}, and \eqref{eq:subset_of_diagonalized_systems_t} via the following conditions
%\begin{equation} \label{eq:final_eigenproblem_formulation}
%\left[ i\sigma \textbf{I} - \textbf{A}\right]\cdot\textbf{v} = 0
%\end{equation}
%where
%\[  
%\textbf{A} =
%\left[
%\begin{BMAT}(rc){c.c.c}{c.c.c}
%%%%%%%%%%%%%%%%%%%%%%%%%%%%%%%%%%
%\textbf{B}_{t} & 0 & 0 \\
%0 & \textbf{B}_{R} & 0 \\
%0 & 0 & \textbf{B}_{i}
%%%%%%%%%%%%%%%%%%%%%%%%%%%%%%%%%%%
%\end{BMAT}
%\right]\]
\begin{subnumcases}{\label{eq:closures_to_stability_problem}}
\hat{U}_{R_0} = 0 \\
\left. \frac{d}{dx} \hat{p}_R \right|_0 = 0 \\
\hat{p}_J = \hat{p}_{R_1} = \hat{p}_{i_0} = \hat{p}_{t_1} \\
\hat{p}_c = \hat{p}_{i_1} = \hat{p}_{t_0} \\
\hat{T}_{t_1} = \frac{1}{\rho_0\,R} \frac{\gamma-1}{\gamma} \hat{p}_J \\
\hat{T}_{t_0} =\frac{1}{\rho_0\,R}  \frac{\gamma-1}{\gamma} \hat{p}_{c}
\end{subnumcases}
representing, respectively, the hard-wall condition on the left end of the resonator, continuity of pressure at the junction and in the compliance, and an isentropic closure for the temperature fluctuations at the two ends of the pulse tube. The final eigenvalue problem can now be built by first combining \eqref{eq:subset_of_diagonalized_systems_R}, \eqref{eq:subset_of_diagonalized_systems_i}, and \eqref{eq:subset_of_diagonalized_systems_t}, into one system of equations,
%\begin{equation}{\label{eq:complete_eigenvalue_problem}}
%\left[ i\sigma \textbf{I} - \textbf{A}\right] \cdot \textbf{v} = 0
%\end{equation}
%The matrix $\textbf{A}$ is built by first stacking $\textbf{B}_{t}$, $\textbf{B}_{R}$, and $\textbf{B}_{i}$ along the diagonal, obtaining
\begin{equation} \label{eq:pseudo_final_eigenvalue_problem}
\left( i\sigma \textbf{I} - \left[
\begin{BMAT}(rc){c.c.c}{c.c.c}
%%%%%%%%%%%%%%%%%%%%%%%%%%%%%%%%%
\textbf{B}_{t} & 0 & 0 \\
0 & \textbf{B}_{R} & 0 \\
0 & 0 & \textbf{B}_{i}
%%%%%%%%%%%%%%%%%%%%%%%%%%%%%%%%%%
\end{BMAT}
\right] \right) \cdot \textbf{v} = 0,
\end{equation}
where $\textbf{v} = \{ \textbf{u}_t ; \textbf{u}_R ; \textbf{u}_i \}$, and then incorporating the conditions \eqref{eq:closures_to_stability_problem} and the equations \eqref{eq:complianceODE_infreq} and \eqref{eq:junctionODE_infreq} to close the problem. Each of the conditions in \eqref{eq:closures_to_stability_problem}, \eqref{eq:complianceODE_infreq}, \eqref{eq:junctionODE_infreq} replaces one corresponding equation in \eqref{eq:pseudo_final_eigenvalue_problem}, therefore, not affecting the rank of the system. The eigenvalue structure is finally recovered by absorbing the equations that do not contain $\sigma$ (i.e. the ones deriving from \eqref{eq:closures_to_stability_problem}) via Gaussian elimination.

The linear modeling framework composed of equations \eqref{eq:complianceODE_infreq} and \eqref{eq:junctionODE_infreq}, \eqref{eq:linearEuler_isentropic}, and \eqref{eq:linearEuler_nonisentropic} is first tested against the three-dimensional numerical simulation data for verification purposes. First, the acoustic impedance at different axial positions in the resonator, obtained by numerically integrating \eqref{eq:linearEuler_isentropic}, has been quantitatively verified against the simulation data, as well as the constants $w_{c_0}$ in \eqref{eq:complianceODE_infreq} and $w_{J_0}$ in \eqref{eq:junctionODE_infreq}, which are equal to 135$\times$10$^6$ N/m$^5$ and 312$\times$10$^6$ N/m$^5$, respectively. Good agreement is also obtained by directly integrating \eqref{eq:linearEuler_nonisentropic} from section $t_0$ to $t_1$ and $i_0$ to $i_1$ (figure \ref{fig:linearStabilityAnalysis}) using data from the numerical simulations as initial conditions (table \ref{tbl:phase_and_amplitude_data}). The integration has been carried out with the exact value of the frequency and growth rate extracted from the simulations (figure \ref{fig:growthRateAndLimitCycleCompare}). Numerical trials have shown that, for a given i.c. and base state, the direct integration of \eqref{eq:linearEuler_nonisentropic} is much more sensitive to the angular frequency, $\omega$ than in the growth rate $\alpha$, which suggests that the prediction of the latter, in thermoacoustic systems, is potentially problematic, especially within the framework of linear modeling in the spectral domain.

\begin{table}
  \centering
  \caption{Comparison between linear theory and simulation data for $T_h=460$K and $T_h=500$K extracted at locations shown in figure \ref{fig:linearStabilityAnalysis} at $t=0.55$ s. The linearized equations \eqref{eq:linearEuler_nonisentropic} are integrated from $t_0$ to $t_1$ and $i_0$ to $i_1$ (figure \ref{fig:linearStabilityAnalysis}) with initial conditions, at $t_0$ and $i_0$, respectively, and $\alpha$, $\omega$, $T_0(x)$ and $\rho_0(x)$ taken from the numerical simulations (figures \ref{fig:tempDensity_parcelTracking}a,\ref{fig:growthRateAndLimitCycleCompare}).}
  \label{tbl:phase_and_amplitude_data}
\resizebox{\textwidth}{!}{%
  \begin{tabular}{ccccc}
  \multicolumn{5}{l}{}\\
      & \multicolumn{2}{c}{$T_h=460 K$} & \multicolumn{2}{c}{$T_h=500 K$}\vspace*{0.1cm}\\
      &  simulation data & linear theory & simulation data & linear theory\vspace*{0.1cm}\\
      \cline{2-5}
\multicolumn{5}{l}{}\\
\multicolumn{5}{l}{\emph{pulse tube:}}\vspace*{0.2cm}\\
    $\hat{p}_{t_0}$  & (1035.43 Pa,  0$^\circ{}$)  		  	& used as i.c.					  	    & (1566.54 Pa,  0$^\circ{}$)  		  	  & used as i.c. \\
    $\hat{U}_{t_0}$  & (0.00159 m$^{3}$/s,  -39.0$^\circ{}$)	& used as i.c.				   	    & (0.00229 m$^{3}$/s,  -36.9$^\circ{}$) & used as i.c.\\ 
  $\hat{U}_{t_1}$  & (0.00586 m$^{3}$/s,  -68.6$^\circ{}$)& (0.00651 m$^{3}$/s, -70.2$^\circ{}$) & (0.00895 m$^{3}$/s,  -66.9$^\circ{}$) & (0.01046 m$^{3}$/s, -69.0$^\circ{}$) \\   
  $\hat{p}_{t_1}$  & (952.30 Pa,  3.28$^\circ{}$)			& (955.75 Pa, 3.11$^\circ{}$)		    & (1437.71 Pa,  3.37$^\circ{}$)	         & (1442.83 Pa, 3.03$^\circ{}$) \\   
%    $\hat{U}_{t_1}$  & (0.00636 m$^{3}$/s,  -70.2$^\circ{}$)	& (0.00697 m$^3$/s, -70.1$^\circ{}$)    & (0.00971 m$^{3}$/s,  -68.6$^\circ{}$)  & (0.0110 m$^3$/s, -68.4 $^\circ{}$) \\
%    $\hat{p}_{t_1}$  & (944.92 Pa, 3.17$^\circ{}$)			& (941.7 Pa, 2.95$^\circ{}$)		    & (1426.50 Pa, 3.25$^\circ{}$)               & (1417.34 Pa, 2.78$^\circ{}$) \\
    \multicolumn{5}{l}{}\\
\multicolumn{5}{l}{\emph{inertance:}}\vspace*{0.2cm}\\
    $\hat{U}_{i_0}$  & (0.00349 m$^{3}$/s,  68.8$^\circ{}$)	& used as i.c.					           & (0.00537 m$^{3}$/s,  69.0$^\circ{}$)   & used as i.c. \\
    $\hat{p}_{i_0}$  & (946.04 Pa, 3.01$^\circ{}$)			& used as i.c.				  	           & (1427.41 Pa, 3.09$^\circ{}$)		  & used as i.c. \\    
    $\hat{U}_{i_1}$  & (0.00216 m$^{3}$/s,  55.0$^\circ{}$)	& (0.00218 m$^3$/s, 53.5$^\circ{}$)	    & (0.00337 m$^{3}$/s,  56.2$^\circ{}$)	  & (0.00337 m$^3$/s, 54.2$^\circ{}$) \\
    $\hat{p}_{i_1}$  & (1031.70 Pa,  0.15$^\circ{}$)		& (1024.34 Pa, 0.07$^\circ{}$)              & (1561.24 Pa,  0.17$^\circ{}$)		  & (1563.52 Pa, -0.23$^\circ{}$) \\    %
%    \multicolumn{5}{l}{}\\
%\multicolumn{5}{l}{\emph{junction:}}\vspace*{0.2cm}\\
%    $\hat{U}_R$  &  \multicolumn{2}{c}{(0.00985 m$^{3}$/s, 1.63$^\circ{}$)} &  \multicolumn{2}{c}{(0.01507 m$^{3}$/s,1.64$^\circ{}$)} \\
%    $\hat{U}_d$  &  \multicolumn{2}{c}{(0.00068 m$^{3}$/s, 1.08$^\circ{}$)} &  \multicolumn{2}{c}{(0.00112 m$^{3}$/s,1.06$^\circ{}$)} \\
%    $c_0$  &  \multicolumn{2}{c}{0.106} &  \multicolumn{2}{c}{0.115}\vspace*{0.1cm}\\
  \end{tabular}
  }
 \end{table}

The eigenvalues $\sigma$ are finally calculated by directly solving the eigenvalue problem \eqref{eq:complete_eigenvalue_problem} for operating conditions ranging between $\tau = 1.25$ and $\tau=1.85$, where $\tau = T_h/T_c$ is the hot-to-cold temperature ratio. The segments representing the pulse tube, inertance and resonator were discretized with 256, 64 and 32 points respectively, with a forth-order spatial polynomial reconstruction. The frequency of instability and its variation with $\tau$ are predicted within a $\sim$ 0.2 Hz error (figure \ref{fig:growthRateAndLimitCycleCompare}a). The operating frequency of the system could also be predicted (with a $\sim$ 1 Hz error) by simply solving the eigenvalue problem \eqref{eq:linearEuler_isentropic} in the complete variable area resonator alone (without the pulse tube), in accordance with phase distribution shown in figure \ref{fig:acousticNetwork}b, which is consistent with simple standing wave resonance. The growth rate is slightly over-predicted (figure \ref{fig:growthRateAndLimitCycleCompare}b) having neglected viscous and nonlinear losses. Overall, the quantitative agreement is very encouraging, serving both as a verification step for the full Navier-Stokes calculations and to gain insight into the nature of the instability, also briefly discussed in the following section.

The complete eigenvalue problem could also be solved by iteratively integrating \eqref{eq:linearEuler_nonisentropic} and \eqref{eq:linearEuler_nonisentropic_diagonalized} in space starting from a given set of initial conditions or guesses. This approach is adopted in {\sc DeltaEC} \citep{WardS_JAcouSocAm_1994} to predict the limit-cycle pressure and velocity distributions in TAEs, which is a valid approximation in the case of relatively low pressure amplitudes, limited waveform distortion and simple geometries. In previous numerical trials this approach has proven to be unsuccessful in the context of the present device, especially in predicting the correct growth rate. A fully implicit spatial formulation, similar to a Helmholtz solvers used in reactive flows \citep{PoinsotV_numComb_2011}, which has been adopted in the present context, is the only one that has proven to be robust, cost effective and reliable.

\subsection{Supercritical Hopf Bifurcation} \label{sec:HopfBifurcation}

\begin{figure}
\centering
 \includegraphics[keepaspectratio=true,width=\linewidth]{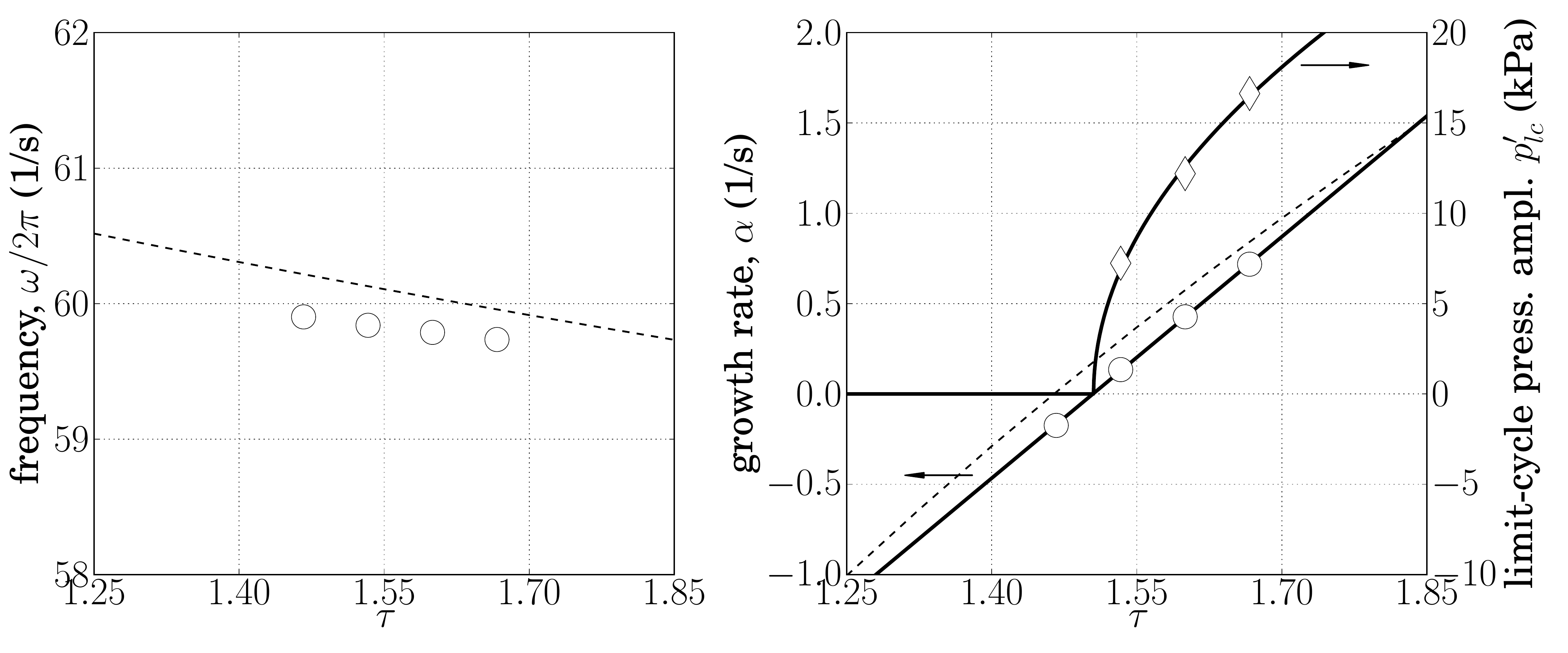}
 \put(-385,2){(a)}
  \put(-15,2){(b)}
 \caption{Frequency, $\omega/2\pi$ (a), growth rate, $\alpha$, and limit-cycle pressure amplitude ${p_{lc}'}$ (b) versus hot-to-cold temperature ratio, $\tau = T_h/T_c$. Linear stability model (-$\,$-$\,$-), numerical simulations (symbols) with corresponding fitting (\textemdash) yielding critical temperature ratio $\tau_{cr}=1.505$ and $\left.p_{lc}'\right|_{\delta \tau=1} = 41,000$ Pa.}
\label{fig:growthRateAndLimitCycleCompare}
\end{figure}

A linear fit of the growth rates, $\alpha$, extracted from the numerical simulation data versus the temperature ratio $\tau = T_h/T_c$ (figure \ref{fig:growthRateAndLimitCycleCompare}b) suggests a critical value of $\tau_{cr} = 1.505$. This is in perfect agreement with the same value obtained by fitting the functional form suggested by the supercritical Hopf bifurcation model
\begin{equation} \label{eq:HopfsqrtScaling}
p'_{lc} = \left.p'_{lc}\right|_{\delta\tau = 1}\sqrt{\tau-\tau_{cr}},
\end{equation}
to the limit-cycle pressure amplitudes $p'_{lc}$ versus $\tau$. A similar result is obtained via nonlinear modeling by \cite{MariappanS_JFM_2011}. The fitting parameters in \eqref{eq:HopfsqrtScaling} are $\left.p'_{lc}\right|_{\delta \tau = 1}$ (dimensional) and $\tau_{cr}$ (dimensionless). Moreover, two assumptions required by the Hopf bifurcation theorem are also satisfied: the non-hyperbolicity condition, $\alpha=0$  and $\omega \ne 0$ at $\tau_{cr}$ (figure \ref{fig:growthRateAndLimitCycleCompare}a), and the transversality condition, d$\alpha$/d$\tau \ne$ 0 at $\tau_{cr}$ (figure \ref{fig:growthRateAndLimitCycleCompare}b). These results have important implication on the parametrization of nonlinear fluxes, discussed later in section \ref{eq:parametrization_nonlinear_fluxes}.

%\Carlo{say that these results are comforting especially given the adoption of ENO. You should also point out that the growth rate might be sensitive to initial conditions that we were not able to control. Despite the consistency with the Hopf bifurcation model }

\section{Nonlinear Regime}  \label{Results::NonlinearRegime}

The analysis carried out so far has been exclusively based on the assumption of linear acoustic perturbations and therefore limited to the start-up phase. Nonlinear effects, however, are already detectable after only a few cycles of operation. These include the departure from exponential growth of the pressure amplitude (figure \ref{fig:TimesSeriesSemiLog}), the presence of broadband fine-scale flow structures associated with transitional turbulence, and a drift in the fluid parcels' velocity, already noticable in the REG/HX during the startup phase (figure \ref{fig:tempDensity_parcelTracking}b). The latter phenomenon is known as acoustic streaming, which is the focus on this section.

The most dramatic manifestation of acoustic streaming in traveling-wave TAEs is the advective heat leakage from the hot heat exchanger (figure \ref{fig:limitcycleMovieframes}) which requires the introduction of a secondary ambient heat exchanger (AHX2) (figure \ref{fig:computationalsetupNijeholt}) to remove the excess heat and achieve a limit cycle. Acoustic streaming occurs everywhere in the engine and its prediction and its suppression is one of the main technological challenges for the design of efficient TAEs.

\subsection{Direct Modeling of Acoustic Streaming} \label{sec:streamXcalculations}

\begin{figure}
\centering
\includegraphics[keepaspectratio=true,width=0.9\linewidth]{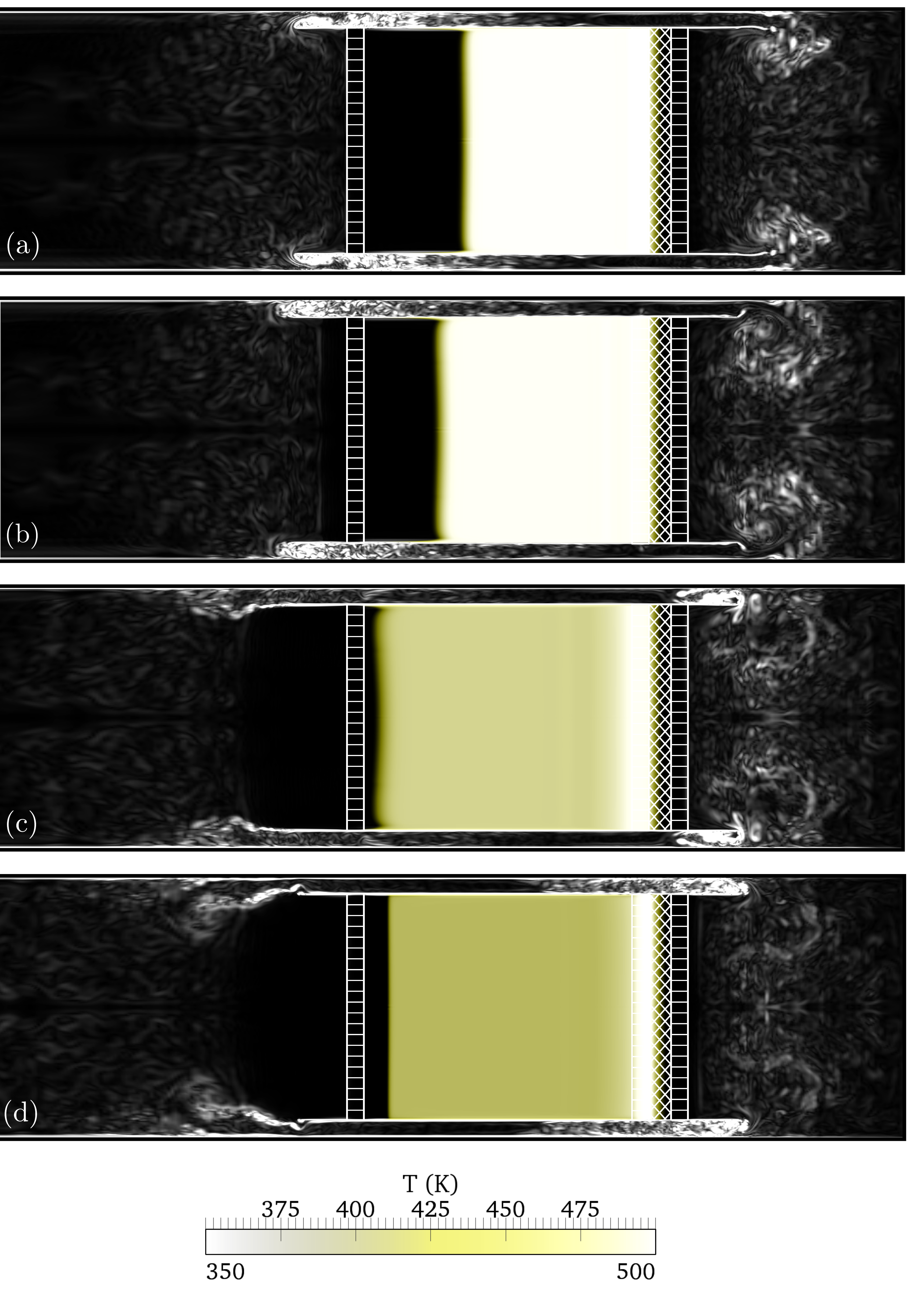}
 \caption{Instantaneous visualizations of temperature contours (colorbar in bottom figure) showing streaming of hot fluid in the thermal buffer tube and vorticity magnitude (white) showing intense vortex shedding and transitional turbulence. Data is shown over one complete acoustic cycle with 90$^\circ$ phase increments from (a) to (d). (Color online)}
\label{fig:limitcycleMovieframes}
\end{figure}

\begin{figure}
  \centering
  \includegraphics[keepaspectratio=true,width=\linewidth]{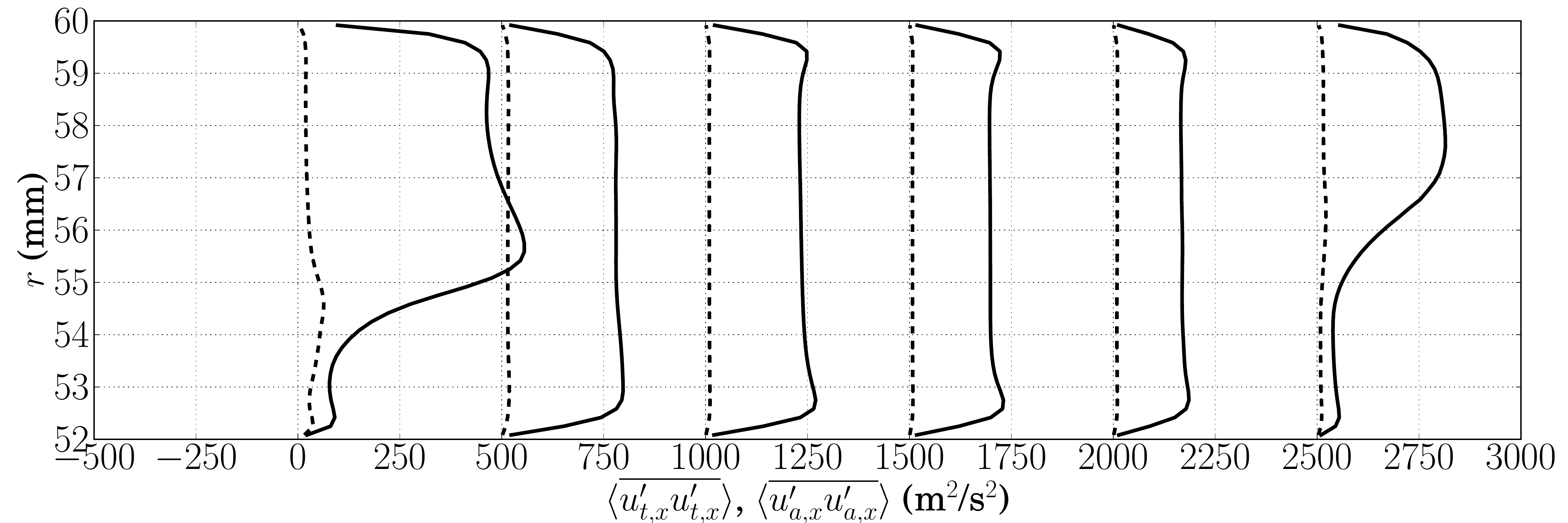}
  \caption{Profiles of cycle- and time-averaged variance of acoustic $\langle \overline{u'_{a,x} u'_{a,x}} \rangle$ (\textemdash) and small-scale $\langle \overline{u'_{t,x} u'_{t,x}} \rangle$ axial velocity fluctuations (\textendash$\,$\textendash) extracted in the inertance, for $x/L_i = 0.01,\,0.2,\,0.40,\,0.60,\,0.80,\,0.99$, from left to right, where $L_i = 0.204 $ m (shifted by 500 m$^2$/s$^2$ for clarity) for $T_h=500$ K and grid C (table \ref{tbl:mesh_temperature_settings}).}
\label{fig:filteredLineData_Turb_vs_Acoustic}
\end{figure}

A triple decomposition can be invoked to separate the streaming flow (rigorously defined below) from the acoustic field and the small-scale high-frequency fluctuations, starting with the Reynolds decomposition
\begin{subnumcases}{\label{eq:solution_decomposition}}
 \rho = \rho_0 + \rho'\\
 p = p_0  + p' \\
 u_i = u_{0,i} + u'
 \end{subnumcases}
where the subscript `$0$' indicates a sharp-spectral-filtered quantity (also used before to indicate mean quantities), such as
\begin{equation} \label{eq:sharpspectral_filter}
u_{0,i} = \overline{u}_i = \int_{-\infty}^{\infty} u_i({\bf x},t+\tau)  \frac{\textrm{sin}(\pi\;f_c\,\tau)}{\pi\;\tau} d\tau
\end{equation}
where $f_c$ is the cut-off frequency. The filtering operation \eqref{eq:sharpspectral_filter} is, in practice, carried out over 6 acoustic periods by adopting Simpson's quadrature rule on the discrete data sampled at 2.2 kHz and $f_c = 0.9\,f$ where $f = \omega/2\pi$ is the acoustic frequency (figure \ref{fig:growthRateAndLimitCycleCompare}a). The remainder of the filtering operation in \eqref{eq:solution_decomposition} can be further decomposed into a purely acoustic (subscript `$a$') and a small-scale component (subscript `$t$'),
\begin{subnumcases}{\label{eq:fluctuation_decomposition}}
\rho' = \rho'_a + \rho'_t \\
p' = p'_a + p'_t\\
u'_i = u'_{a,i}+ u'_{t,i}
 \end{subnumcases}
where the acoustic component can be isolated by applying another filtering operation that removes frequencies higher than $f$ while preserving the full acoustic amplitude. Due to the truncation in time of the filter kernel, this was achieved, in practice, with a cut-off frequency of $f_c > 10\,f$.

Vorticity contours from instantaneous visualizations (figure \ref{fig:limitcycleMovieframes}) suggest that the most intense small-scale fluctuations occur in the feedback inertance. The unsteadiness of the (larger-scale) acoustic fluctuations does not allow turbulence to reach a fully (or even partially) developed state. The Reynolds number based on the Stokes thickness $\delta_\nu = \sqrt{2\nu/\omega}$ and the maximum velocity amplitude at the center of the feedback inertance is approximately $Re_{\delta_\nu} = 480$ (disturbed laminar regime, \cite{Jensen1989JFM}), suggesting that the turbulent kinetic energy generated from the break-up of the vortices rolling-up from the edges of the annular tube is not sustained. Moreover, turbulent stresses extracted for the $T_h=500$ K case (highest drive ratio) and for the finest grid available (table \ref{tbl:mesh_temperature_settings}) are approximately two orders of magnitude smaller than the acoustic stresses (figure \ref{fig:filteredLineData_Turb_vs_Acoustic}) and will be neglected in the following analysis. This choice also accommodates the need to devise a simple predictive modeling framework for the streaming velocity \eqref{eq:streaming_velocity}, which is discussed in the following.

\begin{figure}
\centering
\includegraphics[keepaspectratio=true,width=\linewidth]{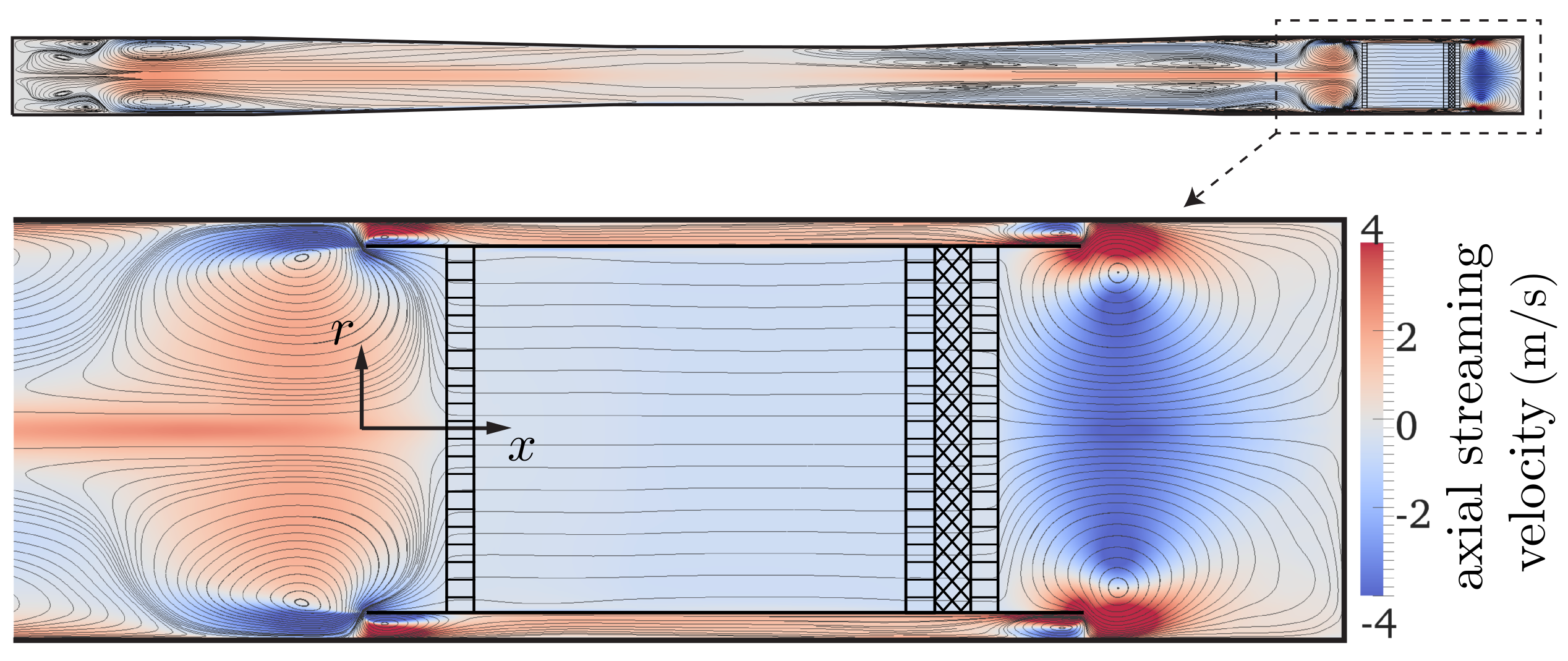}
\put(-15,115){(a)}
\put(-15,0){(b)}
\caption{Contours of axial component of time-averaged streaming velocity $\langle \tilde{u}_x \rangle$ \eqref{eq:streaming_velocity} for $T_h$ = 500K and grid C (table \ref{tbl:mesh_temperature_settings}). Full-scale visualization (a) and zoom on the right end (b). Results have been mirrored about the centerline and streamlines have been only numerically approximated for illustrative purposes. (Color online)}
\label{fig:fullpage_Streaming}
\end{figure}

% \subsection{Modeling of Acoustic Streaming}

Substituting the decomposition \eqref{eq:solution_decomposition} into the time-filtered conservation of mass, ignoring temporal and spatial variations of the filtered density field (a strong assumption, particularly for the regions around the sharp edges and in the thermal buffer tube), assuming that second-order quantities in the small-scale fluctuations are negligible with respect to their acoustic counterpart (figure \ref{fig:filteredLineData_Turb_vs_Acoustic}),
\begin{eqnarray}
 \overline{u_{t,i}' u_{t,i}'} <<  \overline{u_{a,i}' u_{a,i}'} \\
 \overline{\rho_t' u_{t,i}'}  << \overline{\rho_a' u_{a,i}'},
\end{eqnarray}
(despite both being nominally second order) yields the divergence-free condition 
\begin{equation} \label{eq:divergence_of_streaming_velocity}
 \frac{\partial  \tilde{u}_i }{\partial x_i}= 0,
\end{equation}
where $\tilde{u}_i $ is the density-weighted velocity field, $\tilde{u}_i = \overline{\rho u_i}/\rho_0$, defined based on the filtering operation \eqref{eq:sharpspectral_filter}, which, under the assumptions made, can be expressed as
\begin{equation} \label{eq:streaming_velocity}
\tilde{u}_i \simeq u_{0,i} + \frac{\overline{\rho'_a\;u'_{a_i}}}{\rho_0}.
\end{equation}
In the present manuscript the density-weighted velocity field, $\tilde{u}_i$, is adopted as the definition of the streaming velocity based on \eqref{eq:streaming_velocity}, which is second-order accurate in wave amplitude.

The streaming velocity field in our case is axially-symmetric and spans the full extent of the engine (figure \ref{fig:fullpage_Streaming}). Very large and elongated recirculations, of the order of a quarter of the acoustic wavelength, are visible in the resonator and are driven by the wall-normal gradient of the wave-induced shear stresses (not shown). Large recirculations of the order of the resonator radius near the sharp edges of the pulse tube and a mean flow circulating around the pulse tube (following the direction of the amplified waves) are also observed. The latter is called Gedeon (or DC) streaming \citep{Gedeon_1997_Cryocoolers} and is responsible for the advective heat-leakage of the type shown in figure \ref{fig:limitcycleMovieframes}, which limits the efficiency of most traveling-wave thermoacoustic engines (as also discussed later in the context of the present engine).

\begin{figure}
  \centering
  \includegraphics[keepaspectratio=true,width=\linewidth]{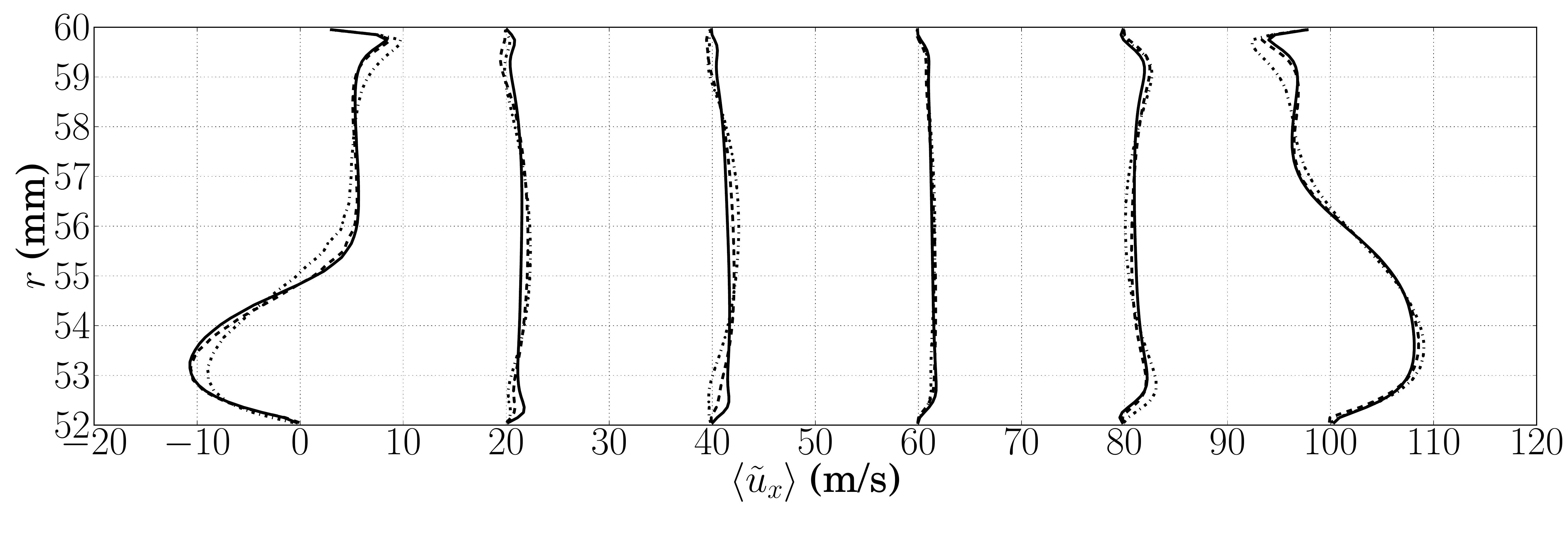}
  \put(-385,20){(a)}
  \vspace*{-0.4cm}
  \includegraphics[keepaspectratio=true,width=\linewidth]{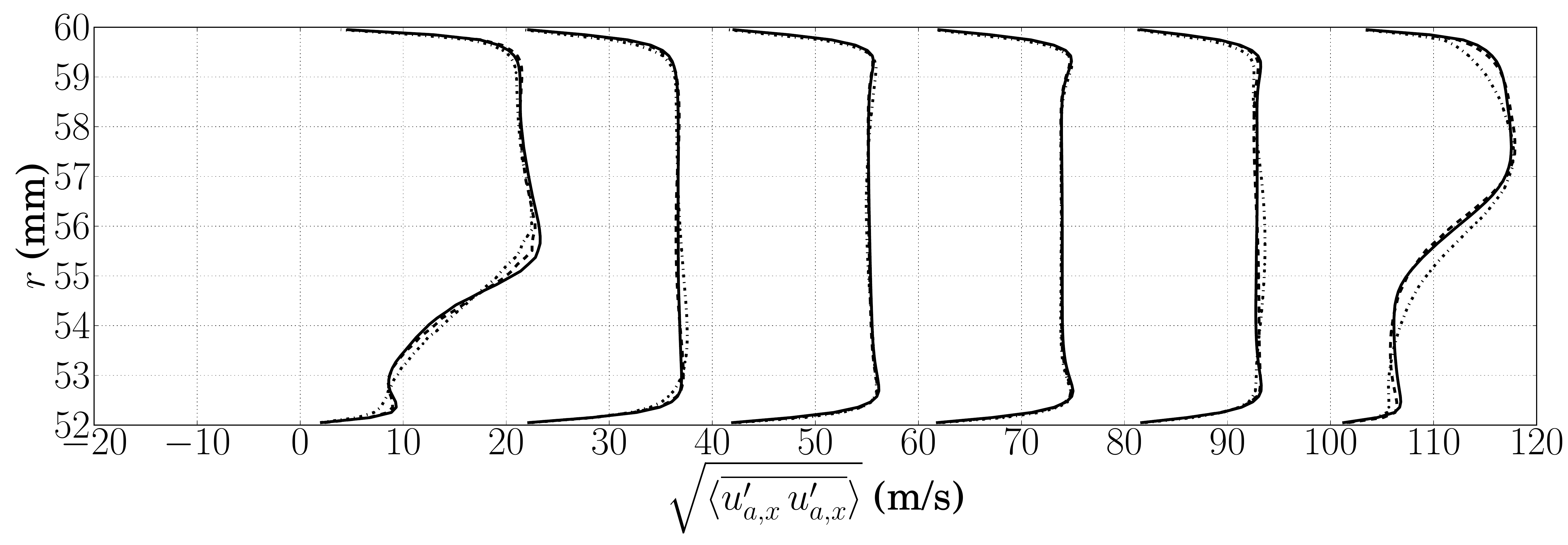}
  \put(-385,20){(b)}
  \vspace*{-0.2cm}
  \includegraphics[keepaspectratio=true,width=\linewidth]{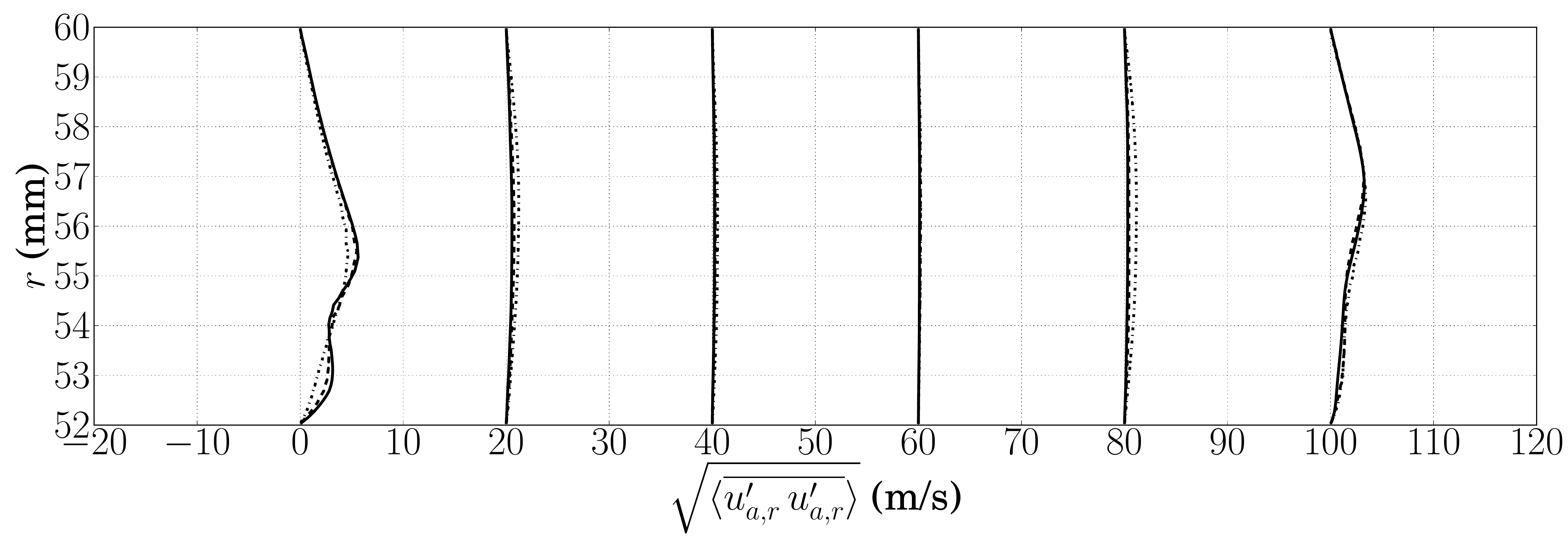}
      \put(-385,20){(c)}
 \vspace*{-0.2cm}
  \includegraphics[keepaspectratio=true,width=\linewidth]{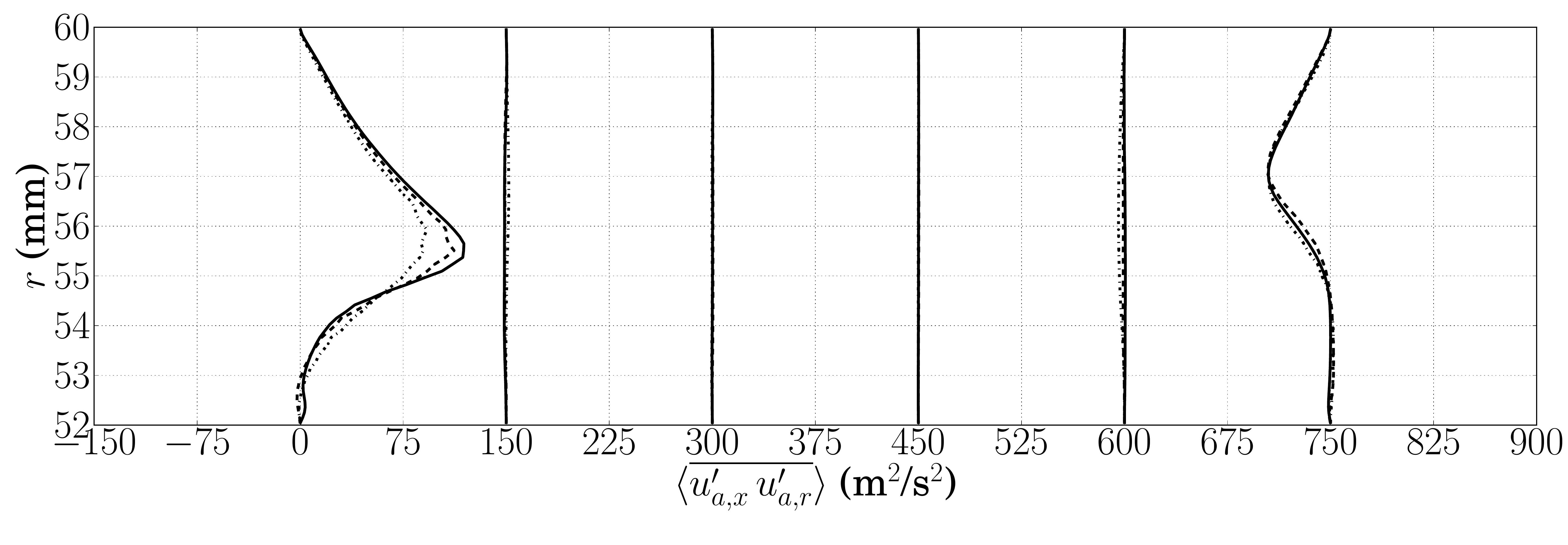}
        \put(-385,20){(d)}
        
  \caption{Profiles of time-averaged streaming velocity (a) and wave-induced Reynolds stresses (b),(c),(d) in the inertance, for $x/L_i = 0.01,\,0.2,\,0.40,\,0.60,\,0.80,\,0.99$, respectively from left to right, where $L_i = 0.204 $ m (shifted by 20 m/s or 150 m$^2$/s$^2$ for clarity). Data for $T_h=500$ K and grid A (\textendash$\,$ $\cdot$ $\,$\textendash), grid B (\textendash$\,$\textendash) and grid C (\textemdash) (table \ref{tbl:mesh_temperature_settings}). The time-averaged volumetric flux through the inertance (intensity of Gedeon streaming) for this drive ratio is approximately 0.0033 m$^3$/s, corresponding to a mean streaming velocity of 1.2 m/s.}
\label{fig:gridconvergence_acousticstresses}
\end{figure}

Assuming that time scales of variation of the filtered quantities $\rho_0$ and $u_{0,i}$ are much longer than the acoustic period, and under the same assumptions underlying the derivation of \eqref{eq:streaming_velocity}, it can be shown that $\tilde{u}_i$ satisfies the incompressible Navier-Stokes equations \citep{RudenkoS_SSS_1977},
\begin{equation} \label{eq:incompressible_NS_streaming}
\frac{\partial \tilde{u}_i}{\partial t} + \frac{\partial }{\partial x_j} \tilde{u}_i\,\tilde{u}_j + \frac{\partial p_0}{\partial x_i} - \nu \nabla^2 \tilde{u}_i =  F_{a,i}
\end{equation}
where $p_0 = P_0/\rho_0$ and the forcing term $F_{a,i}$ is the divergence of the wave-induced Reynolds stresses, which can be expressed to second-order accuracy in wave-amplitude as
\begin{equation} \label{eq:ReynoldsStresses}
 F_{a,i} = - \frac{\partial }{\partial x_j} \overline{u'_{a,j}u'_{a,i}} + \frac{\partial }{\partial x_j} \left\{- \frac{\nu}{\rho_0} \left[ \frac{\partial}{\partial x_j} \overline{\rho'_a u'_{a,i}}+\frac{\partial}{\partial x_i} \overline{\rho'_a u'_{a,j}} \right] + \frac{2}{3} \frac{\nu}{\rho_0} \frac{\partial}{\partial x_k} \overline{\rho'_a u'_{a,k}}\;\delta_{ij} \right\}.
\end{equation}
For large Reynolds numbers based on the streaming velocity, the terms containing the molecular diffusion in \eqref{eq:ReynoldsStresses} can be neglected finally yielding,
\begin{equation} \label{eq:ReynoldsStresses_onlyadv}
 F_{a,i} \simeq - \frac{\partial }{\partial x_j} \overline{u'_{a,j}u'_{a,i}}.
\end{equation}
The maximum value of the stresses $\overline{u'_{a,j}u'_{a,i}}$ is expected to be found in the feedback inertance where the acoustic power is maximum in the system. A grid-sensitivity study on all of the components of the stresses in the inertance (figure \ref{fig:gridconvergence_acousticstresses}) shows monotonic grid-convergence of the stresses from grid A to C (table \ref{tbl:mesh_temperature_settings}).

\begin{figure}
 \centering
\includegraphics[keepaspectratio=true,width=\linewidth]{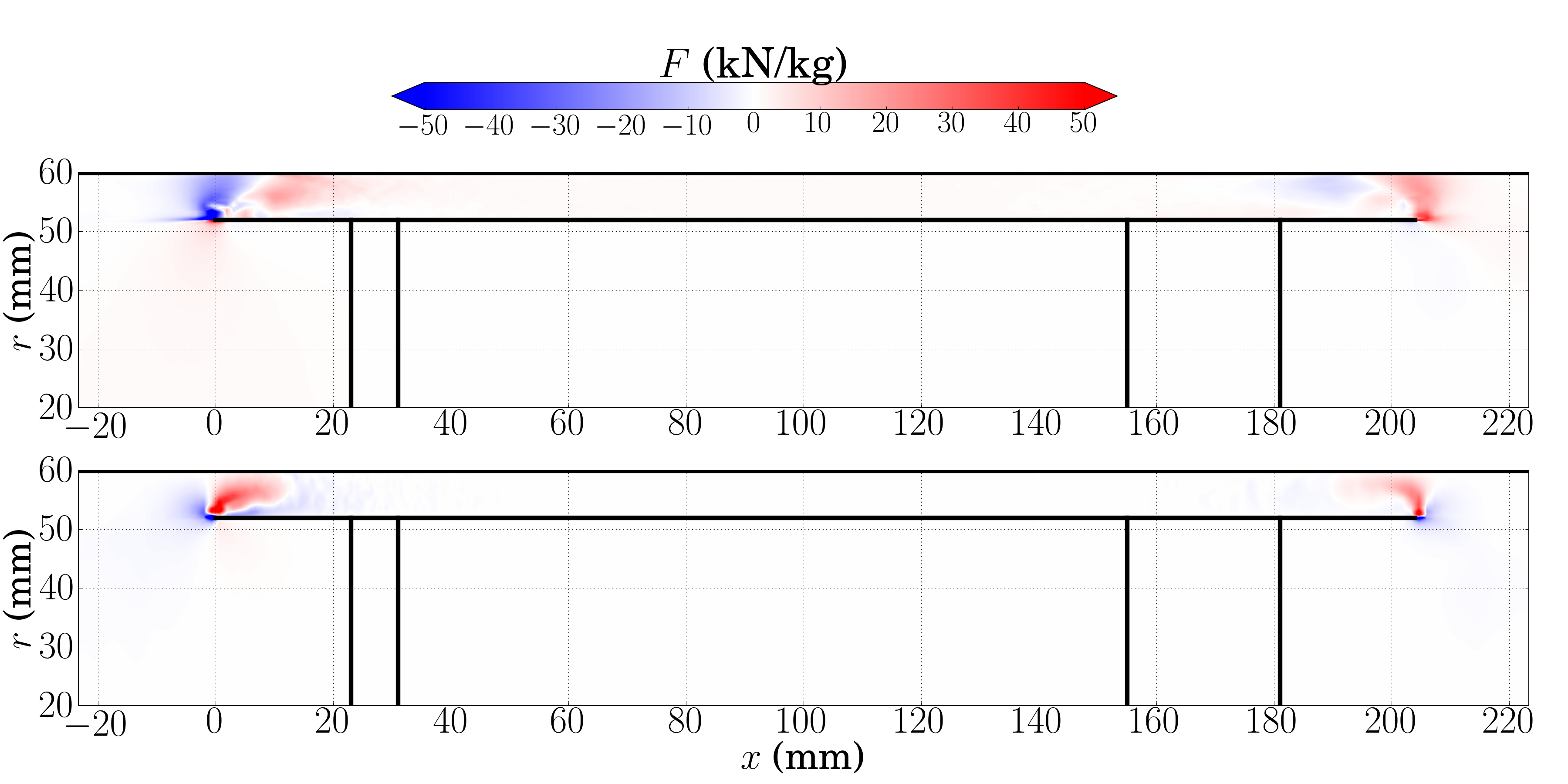}
  \put(-2,82){(a)}
 \put(-2,10){(b)}
\caption{Contour plots of the divergence of wave-induced Reynolds stresses. Axial, $F_{a,x}$ (a) and radial, $F_{a,r}$ (b) components extracted for $T_h=500$K on grid C (table \ref{tbl:mesh_temperature_settings}).}
\label{fig:divergence_wave_induced_stresses}
\end{figure}

The direct evaluation of \eqref{eq:ReynoldsStresses_onlyadv} from the numerical data reveals very high values of $F_{a,i}$ near the sharp edges of the annular tube (figure \ref{fig:divergence_wave_induced_stresses}) which locally drive the large aforementioned recirculations. On the other hand, Gedeon Streaming is driven by the viscous decay of the wave amplitude in the annular inertance. This results in a negative axial gradient of normal stress $\overline{ u_{a,x}' u_{a,x}' }$ visible in both figure \ref{fig:divergence_wave_induced_stresses} and \ref{fig:gridconvergence_acousticstresses}b.

\begin{figure}
\includegraphics[keepaspectratio=true,width=\linewidth]{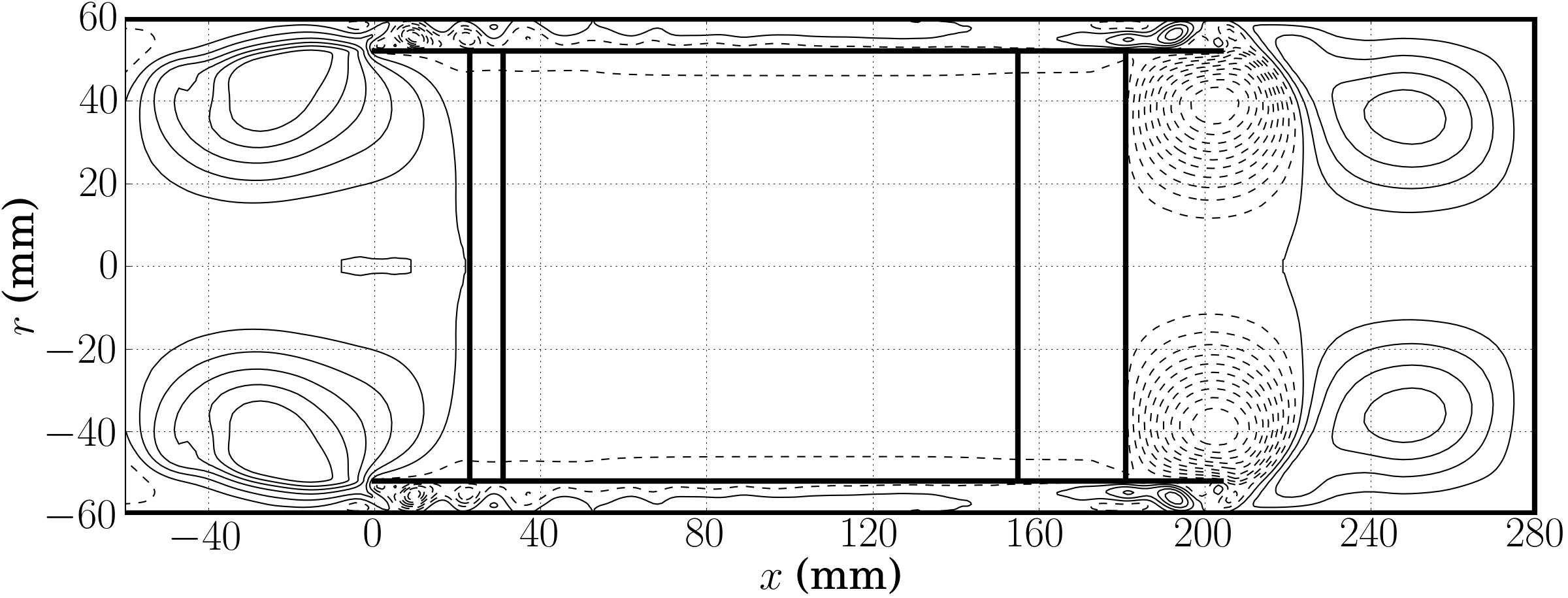}
\caption{Positive (\textemdash) and negative (\textendash$\,$\textendash) iso-levels of Stokes-streamfunction from the incompressible solver \emph{Stream}$^{X}$ driven by the wave-induced Reynolds stresses (figure \ref{fig:divergence_wave_induced_stresses}) extracted from the fully compressible three-dimensional calculations for $T_h=500$K on grid C (table \ref{tbl:mesh_temperature_settings}).}
\label{fig:StreamX_results}
\end{figure}

% Comment on map of divergence of stresses and how it explaines the full_page plot
An axially symmetric numerical model, \emph{Stream$^{X}$} (for more details see Appendix \ref{sec:streamX}), has been developed to directly simulate the streaming velocity field as the solution of the incompressible equations \eqref{eq:incompressible_NS_streaming} driven by the divergence of the wave-induced stresses (figure \ref{fig:divergence_wave_induced_stresses}) extracted from the three-dimensional fully compressible calculations. Secondary features such as steady large-scale recirculations near the sharp edges of the annular tube are only qualitatively reproduced (figure \ref{fig:StreamX_results}) with the appearance of a second recirculation in the compliance, which is not observed in the calculations. The actual target of the present low-order modeling effort is the prediction of the intensity of the Gedeon streaming. In spite of numerical challenges involved in solving incompressible flow in the presence of a sharp edge, the latter is predicted fairly accurately (figure \ref{fig:DCStreaming_and_PulseTubeEnergyScaling}a), especially for low drive ratios. For high drive ratios, resulting in very high limit-cycle acoustic amplitudes, 
the errors associated with the assumptions made in deriving \eqref{eq:streaming_velocity}, \eqref{eq:incompressible_NS_streaming} and \eqref{eq:ReynoldsStresses_onlyadv} become too severe. `Slow' streaming \citep{RudenkoS_SSS_1977} never actually occurs in our engine, where the maximum intensity of $\tilde{u}$ is comparable to the acoustic velocity amplitude in all cases.

\subsection{Efficiency and Energy Fluxes in the Thermal Buffer Tube} \label{eq:parametrization_nonlinear_fluxes}

As the acoustic energy grows during the initial transient, so does the intensity of the Gedeon streaming, increasing the rate of advective transport of hot fluid away from the HHX towards the AHX2. This results in unwanted heat leakage, which lowers the overall energy conversion efficiency. The gradual expansion of the gas in the TBT determines a slow increase of the background pressure in the system, which stops only when the hot temperature front reaches the AHX2. At this point, a rapid increase of the growth rate is observed, as shown by the kink in the time series in figure \ref{fig:TimesSeriesSemiLog}. A limit cycle is only reached later, with a constant background pressure, when the acoustic energy production is balanced by the losses in the system, which include streaming in resonator and dissipation associated with the turbulent vortex shedding from the pulse tube walls.

\begin{figure}
  \centering
  \includegraphics[keepaspectratio=true,width=\linewidth]{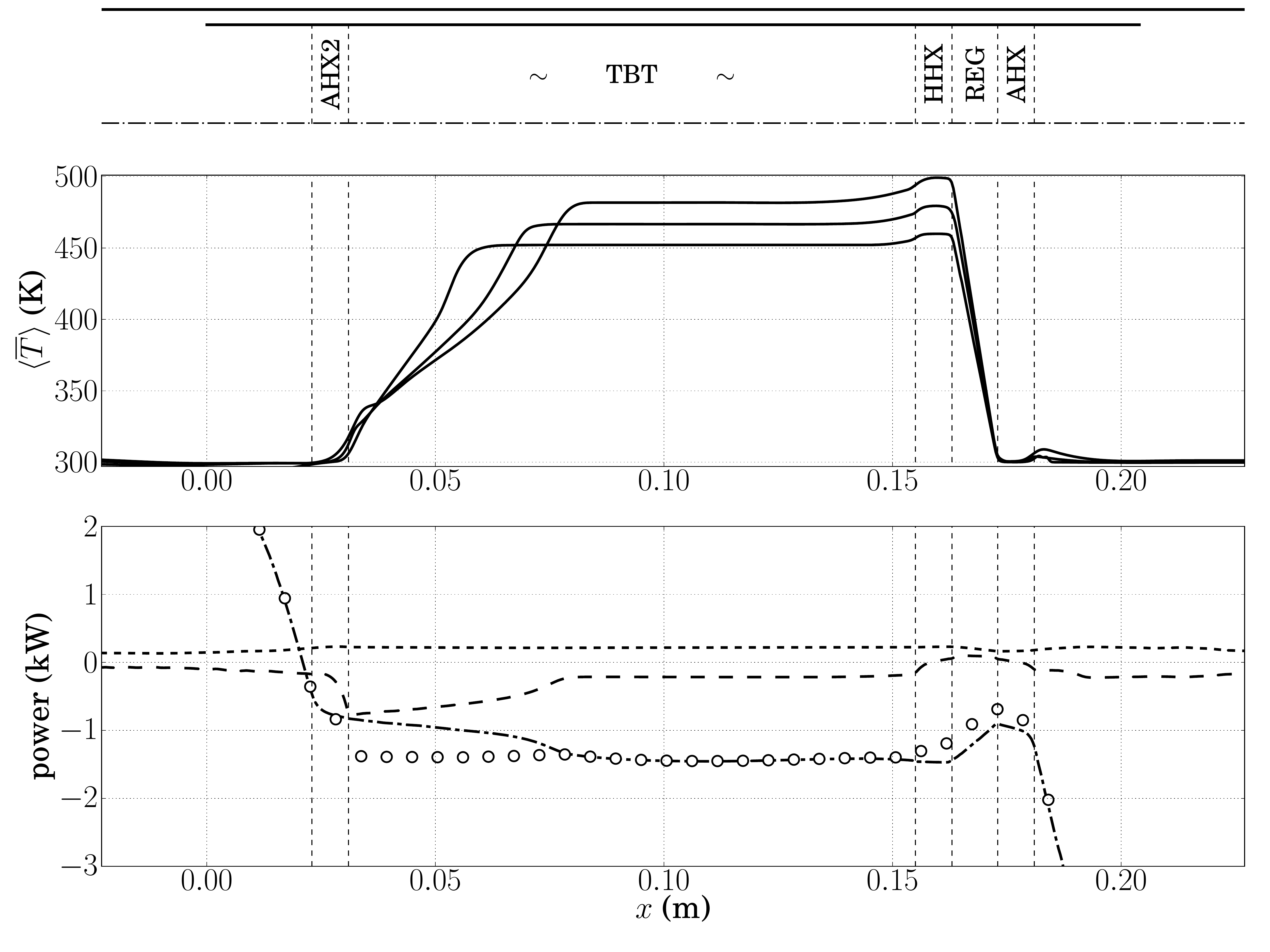}
  \put(-25,160){(a)}
 \put(-25,35){(b)}
  \caption{Axial distribution of mean temperature for $T_h=460$K, $T_h=480$K and $T_h=500$K (a) and surface integrated energy fluxes in \eqref{eq:energy_budget_1D} only for grid C and $T_h=500$ K at limit cycle. Acoustic power (-$\,$-), thermoacoustic heat transport, (\textendash$\,$\textendash), advective heat transport (\textendash$\,$$\cdot$$\,$\textendash), and overall sum ($\circ{}$) (b).}
\label{fig:pulseTubeEnergyBudget}
\end{figure}

The exact conservation equation for the density-averaged internal energy, $\tilde{e} = \overline{\rho\,e}/\overline{\rho}$, reads \citep{Lele_AnnRev_1994}
\begin{equation} \label{eq:energy_budget}
\frac{\partial }{\partial t} \left( \overline{\rho} \tilde{e} \right)  = - \frac{\partial}{\partial x_j} \left( \overline{\rho} \left[ \tilde{u}_j \tilde{e} + \widetilde{h^{''}u^{''}_j} \right] - \overline{q}_j \right) + \frac{\partial}{\partial x_j} \left( \overline{p^{'} u^{''}_j} \right) + \overline{u^{''}_i} \frac{\partial \overline{p}}{\partial x_i} - \overline{p' \frac{\partial u_i^{''}}{\partial x_i}}
\end{equation}
where $u^{''}_j = u_i - \tilde{u}_i$ and $h^{''} = h_i - \tilde{h}$ are the fluctuations of the density-weighted averages of velocity and enthalpy, and $\overline{q}_j$ is the time filtered molecular heat flux. Applying \eqref{eq:energy_budget} to the flow in the TBT approximated as quasi one-dimensional, neglecting small terms and assuming equilibrium conditions, yields
\begin{equation}\label{eq:energy_budget_1D}
\underbrace{ \langle \; \overline{\rho} \tilde{u} \tilde{e} \; \rangle }_\textrm{Advective H.T.}  +  \underbrace{ \langle \; \overline{\rho} \widetilde{h^{''}u^{''}}  \; \rangle }_\textrm{Thermoacoustic H.T.} - \underbrace{ \langle \; \overline{p^{'} u^{''}} \; \rangle }_\textrm{Acoustic Energy Flux}  \simeq  \, \textrm{const}
\end{equation}
which is verified with fairly good approximation in the simulations (figure \ref{fig:pulseTubeEnergyBudget}b). The intensity of the advective heat transport in the TBT is proportional to the mean temperature profile, since the streaming velocity is uniform in this region (figure \ref{fig:fullpage_Streaming}). The adjustment length back to ambient temperature of the mean temperature distribution increases with the drive ratio (figure \ref{fig:pulseTubeEnergyBudget}a). This is due to thermoacoustic heat transport mechanisms. As the hot fluid front is transported with stronger intensity towards the AHX2 by the high-amplitude velocity fluctuations, steeper instantaneous temperature gradients form at the interface between the AHX2 and the TBT (figure \ref{fig:limitcycleMovieframes}). The result is a net cycle-average conductive heat flux in the positive axial direction creating a temperature buffer region. The intensity of the conductive heat flux in \eqref{eq:energy_budget} is, however, neglibile compared to the quantitites in \eqref{eq:energy_budget_1D}, which dominate the energy transport budget in the TBT.

An accurate evaluation of the other terms in \eqref{eq:energy_budget} is, unfortunately, made impractical by the (necessary) application of a second-order ENO reconstruction in the TBT and the smoothing associated with azimuthally averaging the three-dimensional unstructured data. The energy balance expressed by \eqref{eq:energy_budget_1D} is, however, satisfied to a sufficient degree of accuracy to gain insight into the role of the Gedeon streaming in determining the overall efficiency of the device and the scaling of the energy fluxes in \eqref{eq:energy_budget_1D}.

\begin{figure}
  \centering
  \includegraphics[keepaspectratio=true,width=\linewidth]{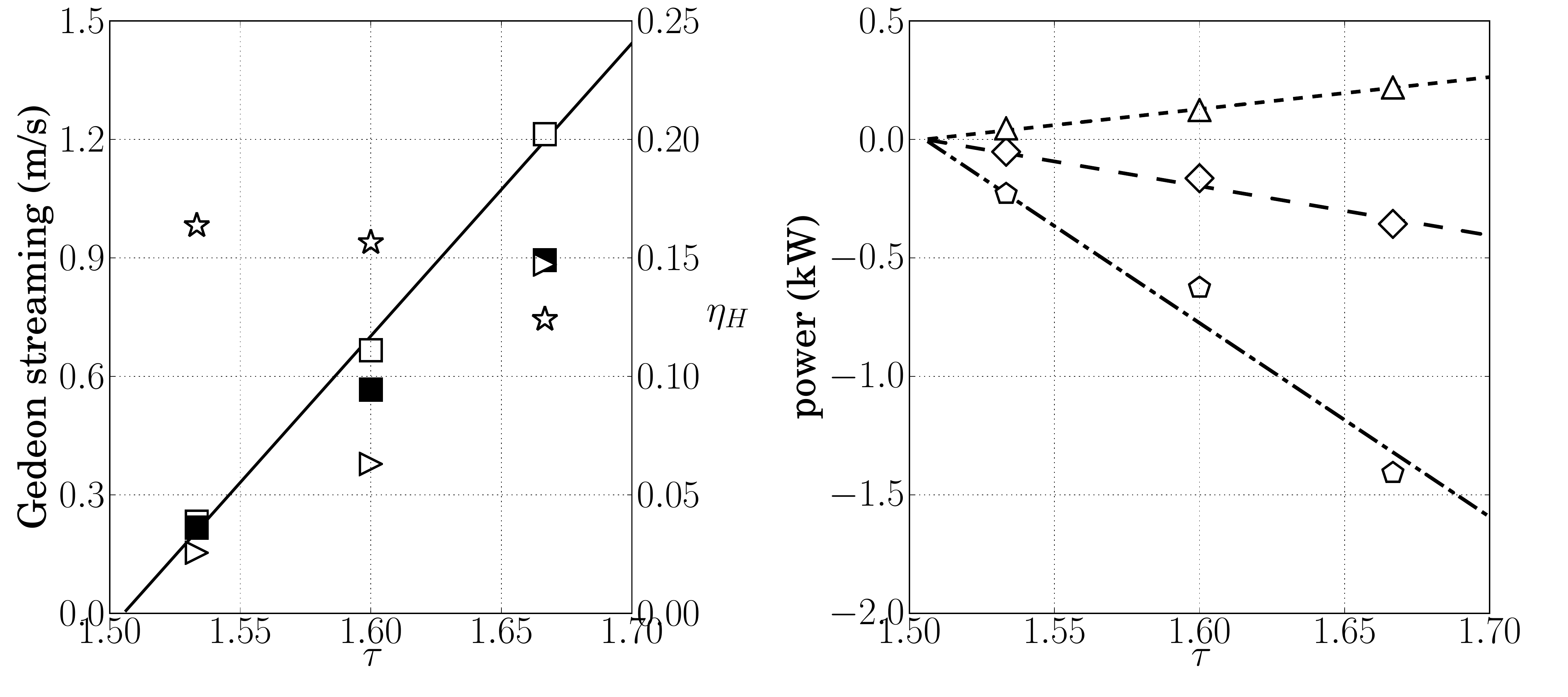}
  \put(-384,20){(a)}
    \put(-10,20){(b)}
  \caption{Intensity of Gedeon streaming versus hot-to-cold temperature ratio for fully compressible Navier-Stokes simulations ($\square$)  with corresponding fit ($\solid$), incompressible model ($\blacksquare$), order-of-magnitude analysis \eqref{eq:veryloworder_DCstreaming} ($\triangleright$), efficiency $\eta_H$ \eqref{eq:efficiency} ($\star$). (a); average nonlinear energy fluxes at limit cycle, full Navier-Stokes simulations (symbols) and corresponding fit based on \eqref{eq:advective_flux_scaling_v2}, \eqref{eq:acoustic_flux_scaling_v2}, \eqref{eq:thermoacoustic_flux_scaling_v2} (lines), respectively for advective heat transport (\pentagon,\textendash$\,$$\cdot$$\,$\textendash), acoustic power ($\bigtriangleup$,-$$-$$-), and thermoacoustic heat transport ({\footnotesize \protect \rotatebox{45}{$\square$}},--$\,$--). }
\label{fig:DCStreaming_and_PulseTubeEnergyScaling} % (b) 
\end{figure}

While no direct energy extraction component (e.g. an acoustic load such as a piezo electric element or a linear alternator) has been included in the setup investigated, a metric for efficiency can still be defined as
\begin{equation}\label{eq:efficiency}
\eta_H = \frac{\langle \; \overline{p^{'} u^{''}} \; \rangle}{ \langle \; \overline{\rho} \widetilde{h^{''}u^{''}}  \; \rangle + \langle \; \overline{\rho} \tilde{u} \tilde{e} \; \rangle},
\end{equation}
which is the ratio between acoustic energy produced and total heat dissipated by the second-ambient heat exchanger. For example, our theoretical device produces $\sim$0.2 kW of acoustic power at $T_h = 500$ K while losing $\sim$2.0 kW mostly due to mean advection caused by Gedeon streaming, achieving a modest efficiency ($\sim\,10$\%). Realistic traveling-wave thermoacoustic engines can reach overall efficiencies of $>20\%$, or even $>30\%$ if built in a cascaded configuration \citep{GardnerS_JASA_2003}. The efficiency directly evaluated from the simulation data (figure \ref{fig:DCStreaming_and_PulseTubeEnergyScaling}a) decreases rapidly with the temperature ratio, which could be expected in thermoacoustic engines with excessive Gedeon streaming controlling the energy balance in the TBT (G. Swift, pers. comm., 2014). However, the uncertainties in the integral quantities in figure \ref{fig:DCStreaming_and_PulseTubeEnergyScaling}b, due to averaging over a limited number (approximately 25) of acoustic cycles, makes the direct metric for the efficiency \eqref{eq:efficiency} not very reliable.

\begin{figure}

  \centering

\includegraphics[keepaspectratio=true,width=0.8\linewidth]{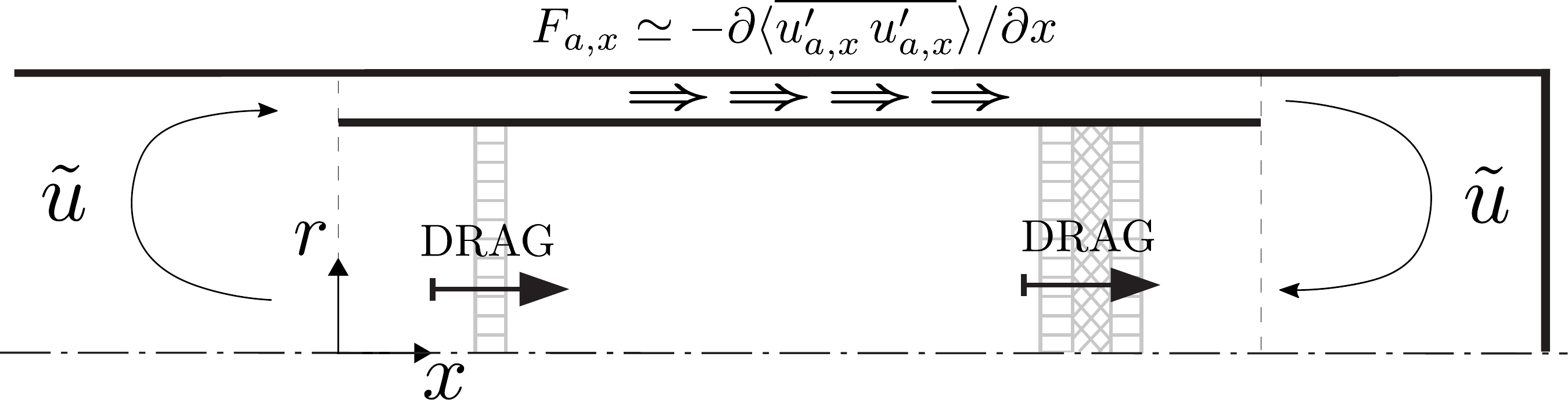}

\includegraphics[keepaspectratio=true,width=0.9\linewidth]{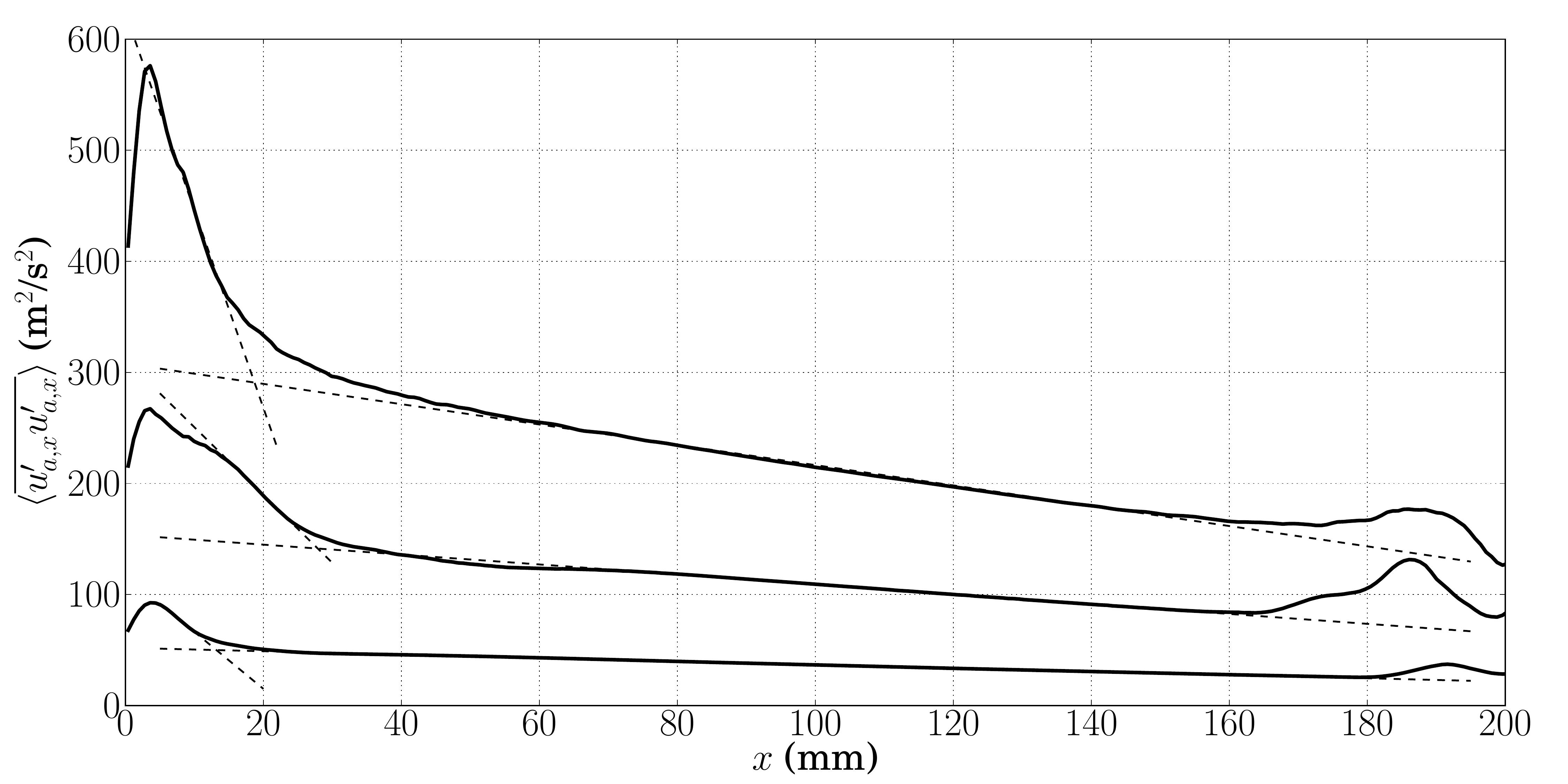}
\put(-150,75){(c)}
\put(-220,52){(b)}
\put(-270,34){(a)}
\caption{Illustration of balance \eqref{eq:veryloworder_DCstreaming} between traveling-wave decay in feedback inertance and drag due to heat-exchangers and regenerator (top). Streamwise profile of axial wave-induced stresses extracted at $r=56$ mm for $T_h=460$K (a), $T_h=480$K (b), and $T_h=500$K (c) with linear fitting (-$\,$-) approximating the mean decay rate.}
\label{fig:StreamingForce_scaling}
\end{figure}

A more robust estimate for $\eta_H$ can be derived by investigating the scaling of the volume-averaged energy fluxes in \eqref{eq:energy_budget_1D} (figure \ref{fig:DCStreaming_and_PulseTubeEnergyScaling}b) with a simple order-of-magnitude analysis and curve fitting. The advective flux in the TBT, driven by the Gedeon streaming, is expected to scale as
\begin{equation}\label{eq:advective_flux_scaling_v0}
 \langle \overline{\rho} \tilde{u} \tilde{e}  \rangle \sim \rho_{0,tbt}  u_{dc} C_v T_h,
\end{equation}
where $\rho_{0,tbt} \sim p_0/(R\,T_c\,\tau)$ is the average density in the TBT and the hot temperature is simply $T_h = T_c\,\tau$. The analytical expression of the Stokes drift due to a freely propagating traveling wave suggests that the intensity of the streaming velocity scales as
\begin{equation} \label{eq:udc_scaling_v0}
u_{dc} \sim \frac{1}{2}\frac{|u'_{lc}|^2}{a_0},
\end{equation}
where $u'_{lc} \sim p'_{lc}/(\rho_0\,a_0)$ is the limit-cycle velocity amplitude. While the quadratic scaling of the streaming velocity is an expected result, a more quantitative estimate for $u_{dc}$ can be obtained by further simplifying the analysis in section \ref{sec:streamXcalculations}, where the streaming velocity is directly modeled based on an hydrodynamic analogy. In fact, by roughly measuring the average spatial decay rate of the axial acoustic stresses in the inertance, $ \langle F_{a,x} \rangle_i$ (figure \ref{fig:StreamingForce_scaling},top), and equating that to the linearized viscous losses in the pulse tube (see (\ref{eq:sourceTerms}a)), the intensity of the Gedeon streaming can be estimated as
\begin{equation}\label{eq:veryloworder_DCstreaming}
u_{dc} \sim \frac{\rho_0\,A_i\,L_i \langle F_{a,x} \rangle_i}{V_\textrm{AHX2}\,R_{c,\textrm{AHX2}} + V_\textrm{HHX}\,R_{c ,\textrm{HHX}} + V_\textrm{REG}\,R_{c,\textrm{REG}} + V_\textrm{AHX}\,R_{c,\textrm{AHX}} },
\end{equation}   
where $V$ and $R_{c}$ are, respectively, the volume and drag coefficients (obtained by linearizing \eqref{eq:sourceTerms}) of the heat-exchanger/regenerators in the pulse tube. The estimate \eqref{eq:veryloworder_DCstreaming} is in good quantitative agreement with the results from section \ref{sec:streamXcalculations} (figure \ref{fig:DCStreaming_and_PulseTubeEnergyScaling}a). The case for $T_h=480$K is an outlier simply due to the lack of grid convergence of the acoustic stresses for this particular case (table \ref{tbl:mesh_temperature_settings}). The quadratic scaling adopted in \eqref{eq:udc_scaling_v0} is, therefore, further justified by \eqref{eq:veryloworder_DCstreaming} where $\langle F_{a,x} \rangle_i \sim {u'_{lc}}^2 $. By invoking the previously derived scaling for the limit-cycle pressure \eqref{eq:HopfsqrtScaling}, \eqref{eq:udc_scaling_v0} becomes
\begin{equation} \label{eq:udc_scaling_v1}
u_{dc} = A_{dc} \frac{1}{2} \frac{1}{\rho_0^2\,a_0^3} \left. {p'_{lc}}^2 \right|_{\delta \tau = 1} \left( \tau - \tau_{cr} \right)
\end{equation}
where the fitting coefficient is $A_{dc}=0.52$ (figure \ref{fig:DCStreaming_and_PulseTubeEnergyScaling}b). Substituting \eqref{eq:udc_scaling_v1} into \eqref{eq:advective_flux_scaling_v0} yields
\begin{equation}\label{eq:advective_flux_scaling_v2}
 \langle \overline{\rho} \tilde{u} \tilde{e}  \rangle = A_{adv} \left[ \frac{C_v\,P_0}{R} A_{dc} \frac{1}{2}\frac{1}{\rho_0^2\,a_0^3} \left.{p'_{lc}}^2\right|_{\delta \tau=1}\right]\left(\tau-\tau_{cr}\right),
\end{equation}
where the fitting coefficient is $A_{adv}=0.51$. The same procedure can be applied to the acoustic energy flux, which scales as
\begin{equation} \label{eq:acoustic_flux_scaling_v1}
\langle\overline{p^{'} u^{''}} \rangle \sim \, p'_{lc} u'_{lc} \frac{1}{2} \textrm{cos}(\Delta \phi)
\end{equation}
where $\Delta \phi$ is the phase difference between pressure and velocity observed in the REG/HX (figure \ref{fig:tempDensity_parcelTracking}c), leading to
\begin{equation} \label{eq:acoustic_flux_scaling_v2}
\langle\overline{p^{'} u^{''}} \rangle = A_{ap}\,\frac{1}{2} \textrm{cos}(\Delta \phi)\, \frac{1}{\rho_0\,a_0}{{\left.{p'_{lc}}^2\right|_{\delta \tau=1}}} \left(\tau-\tau_{cr}\right)
\end{equation}
where the fitting constant is $A_{ap}=0.3$. Finally, the thermoacoustic heat flux is expected to scale as
 \begin{equation}
 \langle\overline{\rho} \widetilde{h^{''}u^{''}}\rangle \sim \rho_{0,tbt} C_p \frac{\partial T}{\partial P} p_{lc}' \frac{p_{lc}'}{\rho_0\,a_0} \frac{1}{2} \textrm{cos}(\Delta \phi)
\end{equation}
where $\Delta \phi$ is needed to account for the correlation between pressure and velocity (like in \eqref{eq:acoustic_flux_scaling_v1}), resulting in
\begin{equation} \label{eq:thermoacoustic_flux_scaling_v2}
 \langle\overline{\rho} \widetilde{h^{''}u^{''}}\rangle = A_{ta} \left[ \frac{1}{2\rho_0\,a_0} \, \left.{p'_{lc}}^2\right|_{\delta \tau=1} \frac{1}{2} \textrm{cos}(\Delta \phi)\right]\left(\tau - \tau_{cr} \right)
\end{equation}
with the fitting coefficient $A_{ta} = 0.93$. With all fluxes scaling as $\sim \left(\tau - \tau_{cr} \right)$, this leads to a robust estimate for the efficiency,
\begin{equation}
\eta_{H,\textrm{avrg}} = 0.13
\end{equation}
effectively averaged over the range of temperature ratios investigated.

Linear acoustic solvers applied at the limit cycle can directly estimate second-order quantities in the acoustic amplitudes such as \eqref{eq:acoustic_flux_scaling_v1} and \eqref{eq:thermoacoustic_flux_scaling_v2}, or even the mean wave-amplitude decay $\langle F_{a,x} \rangle_i$ due to viscous losses in a duct, without prior knowledge of the critical temperature ratio, $\tau_{cr}$. However, process-based parametrizations for the Gedeon streaming, and therefore the advective transport \eqref{eq:advective_flux_scaling_v2}, are still currently missing despite having a first-order impact on the acoustic energy budgets and the efficiency. We have shown that, for slow streaming,commonly found in realistic traveling-wave thermoacoustic engines, a hydrodynamic analogy \citep{Lighthill_JSV_1978} can be invoked and, further simplified leading to a very low-order modeling approach \eqref{eq:veryloworder_DCstreaming}, which is very amenable in the context of simple linear solvers used for engineering prediction of TAEs. Moreover, estimates such as \eqref{eq:veryloworder_DCstreaming} do not require the knowledge of the critical temperature ratio $\tau_{cr}$, making the hydrodynamic analogy investigated in section \ref{sec:streamXcalculations} an attractive modeling paradigm for streaming.

\section{Conclusions}

%% The need to capture complex flow features (induced by the geometrical design) resulted in particularly demanding grid resolution and meshing requirements. In this regard, special care had to be taken to develop load-balancing strategies to ensure maximum performance on large-scale computing platforms.

% What have we done, motivation for the project
We have carried out three-dimensional numerical simulations of a theoretical traveling-wave thermoacoustic heat-engine (TAE). This is the first step in a broader research effort aimed at building multi-fidelity, full-scale prediction tools for TAEs. The goal is to assist technological design by directly simulating, under realistic operating conditions, the physical processes controlling the overall efficiency of such devices. These include the thermoacoustic instability, wave propagation and amplification in the startup phase, the nonlinear effects at the limit cycle (mainly acoustic streaming and turbulence) and the effects of geometrical complexities. The last two are not directly captured in state-of-the-art predictive tools for TAE. 

% How we have picked our computational model
Inspired by the work of \cite{NijeholtTS_2005_JAcoustSA}, we have devised a simple traveling-wave TAE model that could serve as a benchmark case for high-fidelity numerical simulations of similar devices. We have extended such setup to three-dimensions and introduced a second ambient heat-exchanger to achieve a limit cycle, modeling typical fluid dynamic conditions found in thermal buffer tubes. Details omitted in \cite{NijeholtTS_2005_JAcoustSA} regarding the geometry and the modeling of the heat-exchangers and regenerators have been reconstructed to the best of authors' ability and reported in detail. In spite of its theoretical nature, the model retains all the essential critical components, features and complexities of real traveling-wave TAEs.

% What have we discovered, linear regime..
The time integration is carried out from initial quiescent conditions to the limit cycle. It is shown that the mechanisms responsible for the acoustic energy generation and propagation in the system during the start-up phase can be explained with linear acoustics, despite the high amplitude ($\sim$ 1\% of the mean) of the initial perturbation resulting from the activation of the source terms modeling the heat transfer. An analytical linear Lagrangian model shows that the thermoacoustic instability occurring in the regenerator/heat-exchanger (REG/HX) unit intensifies plane waves traveling in the direction of the imposed temperature gradient via a process resembling a thermodynamic Stirling cycle. The result is the establishment of a network of self-amplifying traveling waves looping around the REG/HX unit. A system-wide linear stability model based on Rott's theory accurately predicts the frequency of the (only) unstable mode as well as the the critical temperature ratio, despite not accounting for viscous and other nonlinear losses. The dependency of growth rates and limit-cycle pressure amplitudes on the temperature ratio are shown to be consistent with a supercritical Hopf bifurcation model. No evidence has been found to support subcritical or non-modal instability arguments.

% What have we discovered, nonlinear regime..
At the limit cycle acoustic amplitudes exceed +170dB and nonlinear effects dominate the flow field in the form of transitional turbulence and acoustic streaming. The latter is the occurrence of a quasi-steady flow evolving over time scales much longer than the period of the waves inducing it. The data from the full three-dimensional simulations has allowed to identify the governing processes driving the streaming flow, which are viscous wave amplitude decay in the feedback inertance, periodic vortex ring roll-up and break-up around the sharp edges of the annular tube, and near-wall acoustic shear-stresses in the variable-area resonator.

% Talk about StreamX results (Gedeon Streaming)
An axially-symmetric numerical model based on Stokes-streamfunction formulation has been adopted to directly simulate the streaming flow as the solution of the incompressible Navier-Stokes equations driven by the divergence of the wave-induced Reynolds stresses extracted from the fully compressible three-dimensional calculations. The model correctly reproduces the streaming flow patterns and, in spite of the strong assumptions made and numerical issues associate with geometric singularities, it correctly predicts the intensity of the Gedeon streaming. The latter is responsible for the decrease of the engine's efficiency as the drive ratio is increased, and a robust parametrization for it is warranted. The investigation of the scaling of nonlinear fluxes reveals the importance of prior knowledge of the critical temperature ratio, which may not be straightforwardly achieved by simply relying on linear theory, for more complex systems.

\newpage

\appendix

\section{\emph{Stream}$^X$: an axially symmetric incompressible flow solver model} \label{sec:streamX}

\begin{figure}
 \centering
\includegraphics[keepaspectratio=true,width=.28\linewidth]{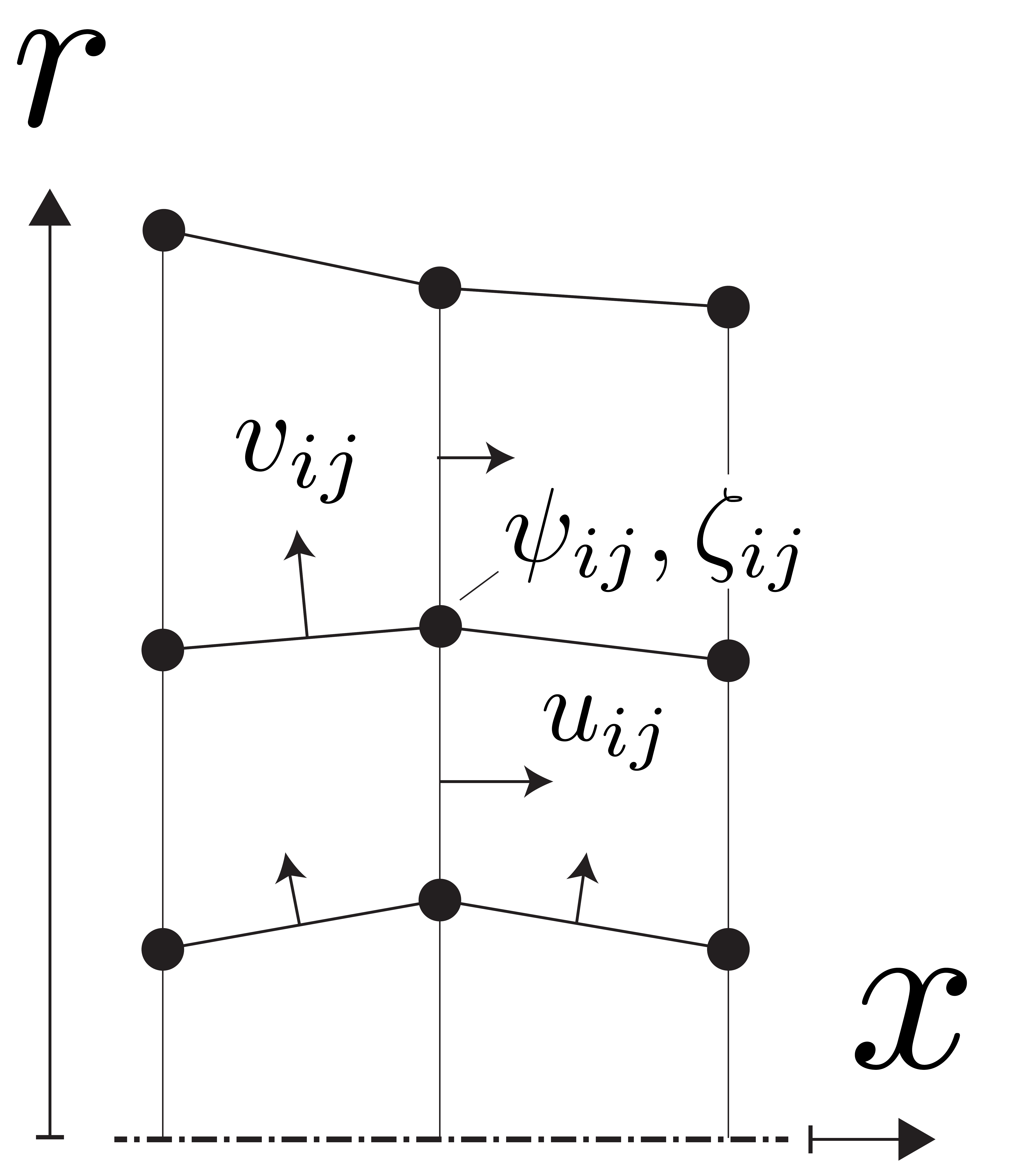}
\put(-120,10){(a)}
\includegraphics[keepaspectratio=true,width=.66\linewidth]{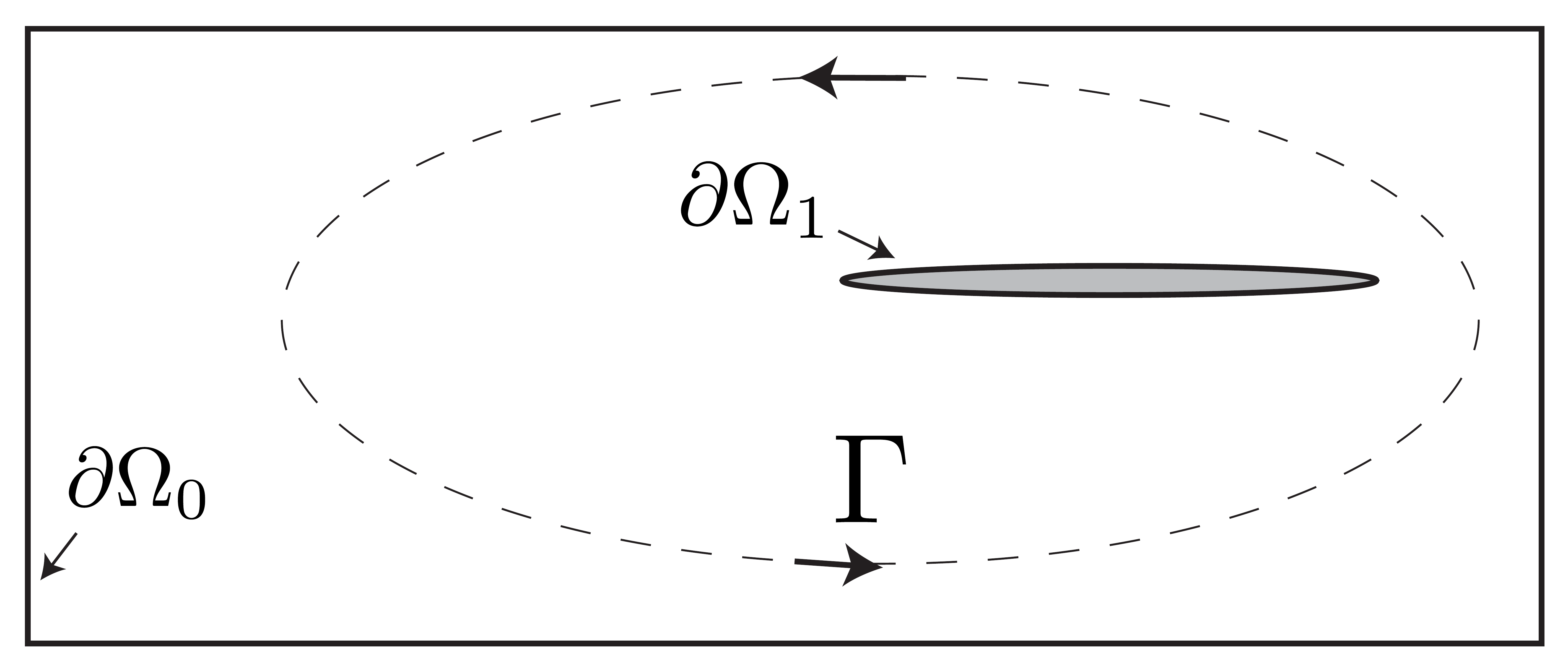}
\put(0,10){(b)}
\caption{Illustration of domain with same degree of connectivity as computational domain in figure \ref{fig:computationalsetupNijeholt} with $\partial \Omega_0$ and $\partial \Omega_1$ representing the resonator walls and the annular tube, respectively, and $\Gamma$ the clock-wise circulation calculated around $\partial \Omega_1$ (a). Variable collocation in \emph{Stream$^{X}$} for axial velocity $u_{ij}$, radial velocity $v_{ij}$, vorticity $\zeta_{ij}$ and Stokes-streamfunction $\Psi_{ij}$ (b). }
\label{fig:StreamX_stuff}
\end{figure}

\emph{Stream$^{X}$} solves the incompressible Navier-Stokes equations in cylindrical coordinates,
\begin{eqnarray}\label{eq:incompressibleNSincylindrical}
&&\frac{\partial u_x}{\partial t} + \frac{1}{r} \frac{\partial }{\partial r} \left(r u_r u_x\right) + \frac{\partial}{\partial x} u_x^2 = - \frac{\partial p}{\partial x} + \nu\left[\frac{1}{r}\frac{\partial }{\partial r} \left( r \frac{\partial u_x}{\partial r} \right) + \frac{\partial^2 u_x}{\partial x^2}\right] + F_x \\
&&\frac{\partial u_r}{\partial t} + \frac{1}{r} \frac{\partial }{\partial r} \left(r\,u_r^2\right) + \frac{\partial}{\partial x} u_x u_r = - \frac{\partial p}{\partial r} + \nu\left[\frac{1}{r}\frac{\partial }{\partial r} \left( r \frac{\partial u_r}{\partial r}  \right) + \frac{\partial^2 u_r}{\partial r^2} - \frac{u_r}{r^2} \right] + F_r
\end{eqnarray}
where $p=P/\rho_0$ and $(F_x,F_r)$ is a body force, relying on a Stokes-streamfunction - vorticity, $\Psi$ - $\zeta$, formulation. A stagghered collocation for $u_x$ and $u_r$, and nodal for $\Psi$ and $\zeta$ is adopted (figure \ref{fig:StreamX_stuff}a). Spatial derivatives are approximated with a second-order central-difference scheme. 

The non-simply connected computational domain (figure \ref{fig:StreamX_stuff}a) requires a special time-advancement strategy to directly solve for $\vec{\Psi}^{n+1} = (0,0,\Psi^{n+1})$, given its unknown value at the boundary $\partial \Omega_1$ (while the value on $\partial \Omega_0$ is fixed and arbitrary). The velocity field is first predicted at the next time step $(u^*,v^*)$ with an explicit Runge Kutta integration of the r.h.s. of \eqref{eq:incompressibleNSincylindrical}, carried out without the pressure terms. This guarantees the exact prediction of vorticity and circulation at time $t^{n+1}$, respectively
\begin{align}\label{eq:info_from_starfield}
\vec{\zeta}^{n+1} =  \vec{\nabla}  \times ({u^*_x},{u^*_r}) \\
\Gamma^{n+1} = \oint ({u^*_x},{u^*_r}) \dot d\vec{l}.
\end{align}
Knowing the exact boundary conditions for $\vec{\Psi}^{n+1}$ on $\partial \Omega_0$ and $\partial \Omega_1$ would allow to directly solve
\begin{equation}
\vec{\nabla}  \times \left( \vec{\nabla} \times \frac{\vec{\Psi}^{n+1}}{r} \right) = \vec{\zeta}^{n+1}
\end{equation}
and complete the time advancement yielding $(u^{n+1}_x,u^{n+1}_r)$. A straightforwad workaround is to express the Stokes-streamfunction at time $t^{n+1}$ as the linear combination
\begin{equation} \label{eq:stokesstreamfunction_lineardecomp}
\Psi^{n+1} = \Psi_A + \alpha\,\Psi_B,
\end{equation}
where $\alpha$ is an unknown coefficient, and $\Psi_A$ and $\Psi_B$ are the solutions to
\begin{align}
&&\vec{\nabla}  \times \left( \vec{\nabla} \times \vec{\Psi}_B/r \right) = 0, \quad \textrm{for} &&\left.\Psi_B\right|_{\partial \Omega_0} = 0, \left.\Psi_B\right|_{\partial \Omega_1} = 1\\
&&\vec{\nabla}  \times \left( \vec{\nabla} \times \vec{\Psi}_A/r \right) = \vec{\zeta}^{n+1}, \quad \textrm{for} &&\left.\Psi_A\right|_{\partial \Omega_0} = 0, \left.\Psi_A\right|_{\partial \Omega_1} = 0
\end{align}
this allows to calculate the circulations $\Gamma_A$ and $\Gamma_B$ and
\begin{equation}
\alpha = (\Gamma - \Gamma_A)/\Gamma_B,
\end{equation}
where $\Gamma$ is known from \eqref{eq:info_from_starfield}. While $\Psi_A$ can be calculated in preprocessing, $\Psi_B$ needs to be re-evaluated at every time step. Finally, $\Psi^{n+1}$ is calculated from \eqref{eq:stokesstreamfunction_lineardecomp}. 

\section*{Acknowledgments}

The authors acknowledge the support of the Precourt Institute for Energy Seed Grant at Stanford and the computational time provided by the NSF-MRI grant on the Stanford Certainty cluster. The authors thank Dr. Gregory Swift for his very useful comments on the draft and acknowledge the help of Prof. Ray Hixon for assisting in the development of the source terms to model the heat-transfer and drag in heat-exchanger and regenerators. Carlo Scalo would like to thank, in particular, Dr. Julien Bodart and Dr. Ivan Bermejo-Moreno for their precious technical help and Jeffrey Lin providing the authors with a complete {\sc DeltaEC} model of the engine, which has lead to the choice of introducing a secondary ambient heat exchanger.

\bibliographystyle{jfm}
\bibliography{references}

\end{document}